\documentclass[prb,showpacs,showkeys]{revtex4-1}
\usepackage{amsfonts}
\usepackage{amsmath}
\usepackage{amssymb}
\usepackage{graphicx}
\usepackage{subfigure}
\begin{document}

\title{First-principles study on oxidation effects in uranium oxides and high-pressure high-temperature behavior of point defects in uranium dioxide}
\author{Hua Y. Geng}
\affiliation{National Key Laboratory of Shock Wave and Detonation
Physics, Institute of Fluid Physics, CAEP,
P.O.Box 919-102 Mianyang, Sichuan 621900, People's Repulic of China}

\author{Hong X. Song}
\affiliation{National Key Laboratory of Shock Wave and Detonation
Physics, Institute of Fluid Physics, CAEP,
P.O.Box 919-102 Mianyang, Sichuan 621900, People's Repulic of China}

\author{K. Jin}
\affiliation{National Key Laboratory of Shock Wave and Detonation
Physics, Institute of Fluid Physics, CAEP,
P.O.Box 919-102 Mianyang, Sichuan 621900, People's Repulic of China}

\author{S. K. Xiang}
\affiliation{National Key Laboratory of Shock Wave and Detonation
Physics, Institute of Fluid Physics, CAEP,
P.O.Box 919-102 Mianyang, Sichuan 621900, People's Repulic of China}

%\author{J. X. Peng}
%\affiliation{National Key Laboratory of Shock Wave and Detonation
%Physics, Institute of Fluid Physics, CAEP;
%P.O.Box 919-102 Mianyang, Sichuan, P. R. China, 621900}

\author{Q. Wu}
\affiliation{National Key Laboratory of Shock Wave and Detonation
Physics, Institute of Fluid Physics, CAEP,
P.O.Box 919-102 Mianyang, Sichuan 621900, People's Repulic of China}


\begin{abstract}
Formation Gibbs free energy of point defects and oxygen clusters in uranium dioxide at high-pressure high-temperature conditions
are calculated from first principles, using the LSDA+\emph{U} approach for
the electronic structure and the Debye model for the lattice vibrations.
The phonon contribution on Frenkel pairs is found to be notable,
whereas it is negligible for the Schottky defect.
Hydrostatic compression changes the formation energies drastically, making
defect concentrations depend more sensitively on pressure. Calculations
show that, if no oxygen clusters are considered, uranium vacancy becomes predominant in overstoichiometric UO$_{2}$
with the aid of the
contribution from lattice vibrations, while compression favors oxygen defects
and suppresses uranium vacancy greatly.
At ambient pressure, however, the experimental observation of predominant oxygen defects
in this regime can be reproduced only
in a form of cuboctahedral clusters, underlining the importance of defect clustering in UO$_{2+x}$.
Making use of the point defect model, an equation of state for non-stoichiometric oxides is established,
which is then applied to describe the shock Hugoniot of UO$_{2+x}$.
Furthermore, the oxidization and compression behavior of uranium monoxide, triuranium octoxide, uranium trioxide, and a series of
defective UO$_{2}$ at zero Kelvin are investigated. The evolution of
mechanical properties and
electronic structures with an
increase of the oxidation degree are analyzed, revealing the transition of the groundstate of
uranium oxides from metallic to Mott insulator and then to charge-transfer insulator due to
the interplay of strongly correlated effects of $5f$ orbitals and the shift of electrons from uranium to oxygen atoms.

\end{abstract}

\volumeyear{year}
\volumenumber{number}
\issuenumber{number}
\eid{identifier}
\maketitle

%\tableofcontents
\section{INTRODUCTION}
\label{sec:intr}

Uranium oxides, especially uranium dioxide (UO$_{2}$), are interesting materials from a fundamental point of view,
not only because of their complex electronic structure arisen from partially filled
5$f$ electron shells
and as materials of great technological importance
in nuclear applications, but also because they are typical nonstoichiometric compounds.
UO$_{2}$ manifests fluorite structure at all temperatures up to the melting point and pressures
up to 40\,GPa,\cite{idiri04,geng07} as well as a wide range of off-stoichiometry.\cite{catlow81,willis64b,willis78}
A comprehensive understanding of its thermodynamical, structural, and kinetic properties
under extreme conditions is very important, especially for nuclear industry since UO$_{2}$ is the main fuel component for pressurized water reactors.
Point defects are the
key contents for these properties, particularly under irradiation as they play a major role in atomic
diffusion. They also act as accommodation of the strong stoichiometry variations that exist in this material
and provide incorporation sites for the fission products.

There are some theoretical studies on the properties of point defects\cite{crocombette01,freyss05,iwasawa06,geng08,nerikar09a,yu09,andersson09b,gupta10,dorado10,crocombette11}
and their incorporation with
fission products\cite{catlow77,grimes91,crocombette02,freyss06,nerikar09,geng2010}
in UO$_{2}$ at
atomic-scale. Early literature used mainly empirical potentials,\cite{catlow77,grimes91}
while recently the ones based on density
functional theory (DFT) prevail. Since the standard local density approximation (LDA) or generalized
gradient approximation (GGA) for the exchange-correlation functional fails to reproduce the fact that
UO$_{2}$ is a Mott-Hubbard insulator with an energy gap split by the strongly correlated
uranium $5f$ electrons,\cite{petit96,dudarev97,dudarev98}
current \emph{state of the art} approaches make use of DFT+\emph{U} formalism or
the hybrid functional method to describe the electronic structure of UO$_{2}$.\cite{geng07,dudarev98,dorado09,hybrid}
With these methods, the formation energy of point defects in UO$_{2}$ at zero Kelvin
were extensively
studied,\cite{iwasawa06,geng08,nerikar09a,yu09,andersson09b,gupta10,dorado10,crocombette11,nerikar09,geng2010}
including both neutral and charged defects. The population of point defects
at finite temperatures and partial oxygen pressures were also investigated based on these 0\,K
formation energies.

After these efforts, understanding of the electronic structure, magnetic ordering,
atomic structural geometry, and energetics
of perfect UO$_{2}$ and single point defects in it was advanced enormously.
Nevertheless a fundamental experiment observation that oxygen defects predominate in the
hyperstoichiometric regime still can not be reproduced, even at a qualitative level.\cite{crocombette01,freyss05,crocombette11}
Only one LSDA+\emph{U} calculation predicted the predominance of oxygen interstitial
in this region.\cite{geng08} However, the agreement is quite marginal.
On the other hand, if taking oxygen defect clustering into account, the predominance of
oxygen defects in a form of cuboctahedral cluster can be easily established,\cite{geng08b,geng08c}
indicating the complicated nature of this problem.

Furthermore, almost all of the previous investigations predicted a negative formation energy
for single oxygen interstitial O$_{i}$, except one of the DFT+\emph{U} calculations with a GGA exchange-correlation functional
that predicted a positive value of 0.1\,eV.\cite{dorado10}
This is sharply in contrast to
other DFT+\emph{U} and empirical results which usually are about $-1\sim-3$\,eV.\cite{geng08,nerikar09a,crocombette11}
It does not sound unreasonable considering that the hyperstoichiometric
phase of UO$_{2}$ is unstable at 0\,K.\cite{higgs07}
However, it is worthwhile to note that a negative formation energy of O$_{i}$
does not conflict with this observation too, since oxygen clusters have a lower energy\cite{geng08,geng08b,geng08c,andersson09} and phase
separation between the ordered phases of fluorite UO$_{2}$ and U$_4$O$_9$\cite{hering52,bevan86,cooper04}
could also lead to the
same phase diagram.
Nevertheless, a formation energy of O$_{i}$ being apparently positive indicates a different oxidization mechanism,
in which the lattice vibrations, or equivalently the phonons, should play a pivotal role at finite temperature and therefore alter the sign
of the formation energy. This part of the contribution to the stability of defects in UO$_{2\pm x}$,
however, was totally absent in almost all of the previous investigations. It now
becomes imperative to assess the role played by this part of free energy on the energetics of defects.

In addition to reducing the free energy, lattice vibration at high temperature also leads to
thermal expansion. The resultant dilation of the system volume inevitably
changes the equilibrium 0\,K formation energy. The influence of this effect on defect populations, as well as that due to possible residual
stress arisen from irradiation damages, are unclear,
which is possibly a partial consequence of the fact that
the hydrostatic compression behavior and
the equation of state (EOS) of solid UO$_{2\pm x}$ drew little attention in the past decades.\cite{yakub06}
These mechanical properties now become important.
It is crucial to know how the fuel components respond
to shock waves generated at nearby explosions to
assess the general safety of nuclear reactors and spent fuel storage facilities.
Attempts to address these problems require a profound understanding of defects
behavior at high-temperature high-pressure conditions, as well as a full
EOS for non-stoichiometric fuel compounds.

In this paper, we evaluate the importance of lattice vibration and hydrostatic compression on
the energetics and population of point defects in UO$_{2}$ with DFT+\emph{U} calculations.
An equation of state model for
off-stoichiometric UO$_{2\pm x}$ is then developed.
It will benefit the understanding of other compounds since UO$_{2}$ is
a typical non-stoichiometric oxide, and an EOS theory developed for it can be easily
adapted for other alike materials.
We also attempt to answer the question of whether oxygen clusters are definitely necessary
or not in order to make oxygen defects the major component
when contributions of phonon and thermal expansion have been taken into account.
In addition, in order to demonstrate the overall trend of oxidation effects and highlight the
crucial role played by UO$_{2\pm x}$, the energetics and electronic structures of uranium
monoxide (UO), triuranium octoxide (U$_{3}$O$_{8}$), and uranium trioxide (UO$_{3}$) are investigated as well.
The calculation methodology and the theoretical formalism
are presented in the next section. The results and discussion are given in
Sec. III, which is then followed by a summary of the main conclusions.

\begin{table}
\caption{\label{tab:coheE} First-principles results for the energy curves of UO, UO$_{2\pm x}$,
$\alpha$-U$_{3}$O$_{8}$, and $\delta$-UO$_{3}$,
where $x$ is the deviation from the stoichiometric composition of uranium dioxide and $N$ is
the total number of atoms in the simulation cell.
$D$ (in eV), $r_{0}$ (in $\mathrm{\AA}$),
and $B_{0}$ (in GPa) are the cohesive energy per atom, the equilibrium lattice parameter of the
effective cubic cell, and the zero pressure bulk modulus, respectively. }
\begin{ruledtabular}
\begin{tabular}{l c c c c c c l} %{10cm}{@{\extracolsep{\fill}}l c c c c c}
  Label & $x$ & $N$ &Functional & $D$ & $r_{0}$ & $B_{0}$ & Phase \\
%            &          &  & &  &\\
\hline
  UO              & $-1$          &8 &GGA-PBE&7.446 &   4.875 & 187.58 & NaCl\,(FM) \\
\hline
  ${}^{u}C4_{1}$  &$-\frac{2}{17}$&49&LSDA+$U$ &8.045&5.513  &198.83 & CaF$_{2}$\,(AFM)\\
  $C4_{-1}$       &$-\frac{1}{16}$&47&LSDA+$U$ &8.171&5.448  &210.35 & CaF$_{2}$\,(AFM)\\
  ${}^{u}C8_{1}$  &$-\frac{2}{33}$&97&LSDA+$U$ &8.135&5.481  &212.39 & CaF$_{2}$\,(AFM)\\
  $C8_{-1}$       &$-\frac{1}{32}$&95&LSDA+$U$ &8.197&5.449 &215.78 &   CaF$_{2}$\,(AFM)\\
  $C1$            &0              &12&LSDA+$U$ &8.219&5.452 & 215.55 &  CaF$_{2}$\,(AFM)\\
  $C8_{1}$        &$\frac{1}{32}$ &97&LSDA+$U$ &8.187&5.443   &221.63  &CaF$_{2}$\,(AFM)\\
  $C4_{1}$        &$\frac{1}{16}$ &49&LSDA+$U$ &8.145&5.439   &202.14  &CaF$_{2}$\,(AFM)\\
  ${}^{u}C8_{-1}$ &$\frac{2}{31}$ &95&LSDA+$U$ &8.154& 5.435  & 212.37 &CaF$_{2}$\,(AFM) \\
  ${}^{u}C4_{-1}$ &$\frac{2}{15}$ &47&LSDA+$U$ &8.085&5.410  &200.38  & CaF$_{2}$\,(AFM)\\
  COT-$o$         &$\frac{5}{32}$ &101&LSDA+$U$ &8.079& 5.436 &210.12 & CaF$_{2}$\,(AFM)\\
  $C1_{1}$        &$\frac{1}{4}$  &13 &LSDA+$U$ &7.937& 5.402 &246.63 & CaF$_{2}$\,(AFM)\\
\hline
  $\alpha$-U$_{3}$O$_{8}$  & $\frac{2}{3}$ &11&GGA-PBE&7.594& 5.537 &165.81 & $Amm2$\,(FM) \\
\hline
  $\delta$-UO$_{3}$  & 1 &4&GGA-PBE& 7.415&4.168 &147.21 & $Pm\overline{3}m$\,(NM) \\
\end{tabular}
\end{ruledtabular}
\end{table}

\section{Methodology}
\subsection{Ground-state energy calculation}
The energetics of UO, U$_{3}$O$_{8}$, UO$_{3}$, and a series of defective UO$_{2}$ in fluorite structures
(to model off-stoichiometric UO$_{2\pm x}$) were
studied through calculating the total energy curves with the
plane-wave method using the density functional theory to treat the electronic
energy as implemented in the VASP code.\cite{vasp,kresse96}
The electron-ion interactions
were described by projector-augmented wave (PAW) pseudopotentials.\cite{blochl94,kresse99}
The $2s^{2}2p^{4}$ electrons in oxygen and $6s^{2}6p^{6}5f^{3}6d^{1}7s^{2}$
in uranium were treated in valence space.
For UO$_{2}$, the electronic exchange-correlation energy was computed by spin-polarized local
density approximation with an
effective on-site Coulomb repulsive interaction to split the partially filled 5\emph{f}
bands that localized on uranium atoms (LSDA+\emph{U}).\cite{anisimov91,liechtenstein95,dudarev98b}
The LSDA+\emph{U} energy functional is given by
\begin{equation}
  E_{\mathrm{LSDA+}U}=E_{\mathrm{LSDA}}+E_{\mathrm{Hub}}-E_{\mathrm{dc}},
\end{equation}
where $E_{\mathrm{LSDA}}$ is the LSDA contribution to the energy, $E_{\mathrm{Hub}}$ is the
electron-electron interactions from the Hubbard term, and $E_{\mathrm{dc}}$ is the double-counting
correction. The LSDA+\emph{U} approximation is thus a correction to the standard LSDA energy functional.
$E_{\mathrm{Hub}}$ and $E_{\mathrm{dc}}$ depend on the occupation matrices of the correlated
orbitals. We used the simplified rotationally invariant approximation for the Hubbard term $E_{\mathrm{Hub}}$
due to Dudarev \emph{et al}.\cite{dudarev98b}
As regards the double counting correction, we used the fully localized limit, whose expression is
\begin{equation}
  E_{\mathrm{dc}}=\frac{U}{2}\hat{N}(\hat{N}-1)-\frac{J}{2}\sum_{\sigma}\hat{N}^{\sigma}(\hat{N}^{\sigma}-1).
\end{equation}
Parameters of the Hubbard
term were taken as $U=4.5$\,eV and $J=0.51$\,eV, which have been checked carefully
for \emph{fluorite} UO$_{2}$.\cite{dudarev00,dudarev97,dudarev98,geng07}
The possible metastable states introduced by strong correlation effects were removed
via a quasiannealing procedure,\cite{geng2010} which has the capability of yielding the same results as an alternative but powerful
approach of monitoring the occupation matrices.\cite{geng2010,dorado-cmt}
Previous investigation implied that the value of the Hubbard parameters depends on
the local coordination environment,\cite{geng07} and thus it should not be the same for different oxides.
Since there are no appropriate values of $U$ and $J$
available for UO, U$_{3}$O$_{8}$, and UO$_{3}$, we used the pure GGA with the
Perdew-Burke-Ernzerhof (PBE) exchange-correlation functional\cite{pbe96}
instead of DFT+\emph{U}, which usually produces
adequate energetics for uranium oxides even though the predicted electronic structure
might be wrong.\cite{geng07}
The validity of such treatment is guaranteed by comparing the calculated lattice parameters
with experimental data.
The cutoff for kinetic energy of plane waves was set as high as 500\,eV to
converge the electronic energy within a few $m$eV per atom.
Integrations over reciprocal space were performed in the irreducible
Brillouin zone with about 8$\sim$36 non-equivalent k-points,
depending on the system size.

In this work, we did not take spin-orbit coupling into account. Previous calculations on metallic
$\alpha$-U,\cite{soderlind02,freyss05} perfect UO$_{2}$,\cite{geng07} and U$_{3}$O$_{8}$\cite{yun11}
indicated that spin-orbit coupling has little effect on the structure and energy variation of uranium oxides, although it is
well known that it is remarkable for magnetic properties. The only exception was reported by Yu \emph{et al.}\cite{yu09}
Because inclusion of spin-orbit coupling in DFT+\emph{U} deteriorates the already notorious problem of metastable states and convergence,
and they did not employ any special technique to tackle this problem, it is highly possible that the reported
results are due to premature calculations that trapped in metastable states. This can be inferred from the jagged variation
(which is quite unphysical) of the calculated lattice
constant with \emph{U}, as shown in Fig.3 of Ref.[\onlinecite{yu09}]. All of this indicates
that their conclusion is still inconclusive. Based on these considerations, we neglected spin-orbit
coupling in all of the following calculations.

The \emph{supercell} method has been employed with periodic boundary conditions.
The arrangement of the structures is listed in Table \ref{tab:coheE}, in which $Cm_{n}$ denotes
a defective UO$_{2}$ structure consisting of $m$ fluorite cubic cells with $n$ oxygen interstitials (or vacancies if $n$ is negative).\cite{geng08}
The superscript $u$ indicates the defect is with respect to uranium. For example,
configuration $^{u}C8_{1}$ has the same geometry as $C8_{1}$ but replacing
the interstitial oxygen with a uranium atom, and $^{u}C8_{-1}$ or $C8_{-1}$
corresponds to removing one lattice atom (uranium and oxygen, respectively) from a structure with 8 fluorite cubic cells ($2\times2\times2$).
Before calculating the energy, all structures were fully relaxed until residual Hellman-Feynman forces (stress) were less than $0.01$\,eV/\AA.

Cohesive energies at different volumes
are extracted from the total energies
by subtracting the contribution of isolated spin-polarized atoms.
They are then fitted
to a Birch-Murnaghan equation of state,
\begin{equation}
  E_{c}(V)=-D+\frac{9}{8}B_{0}V_{0}\left[\left(\frac{V_{0}}{V}\right)^{\frac{2}{3}}-1\right]^{2},
\end{equation}
to facilitate post-analysis, where the equilibrium atomic volume $V_{0}$ can be determined from the effective lattice parameter as listed
in Table \ref{tab:coheE}. With this equation, the pressure at 0\,K can be expressed as
\begin{equation}
  P_{c}(V)=\frac{3}{2}B_{0}\left[\left(\frac{V_{0}}{V}\right)^{\frac{2}{3}}-1\right]\left(\frac{V_{0}}{V}\right)^{\frac{5}{3}}.
\end{equation}

\subsection{Formation Gibbs free energy}

Phonon contribution to free energy and thermodynamic properties can be computed via lattice dynamics or molecular dynamics simulations.
After getting the vibrational density of states $g(\omega)$, the thermal part of the Helmholtz
free energy per atom in quasi-harmonic approximation reads
\begin{equation}
  F_{ph}(V,T)=k_{B}T\int_{0}^{\infty}g(\omega)\ln\left[2\sinh\left(\hbar\omega/2k_{B}T\right)\right]\,d\omega.
  \label{eq:fvib}
\end{equation}
This method is accurate when anharmonic effect is insignificant. But the calculation is extremely
demanding for \emph{ab initio} simulations, especially for defective structures containing more than 90\, atoms without any symmetry,
as discussed here. Crude approximation includes computing the inter-atomic forces from empirical
potentials\cite{watanabe09} instead of by DFT calculations, while for off-stoichiometric compounds it is difficult
to construct an accurate inter-atomic potential that maintains the charge neutrality of the cell generally.

For practical purposes, we resorted to the Debye model, which is less accurate but much easier
for computation. More important is that it provides a convenient description of lattice dynamics
with a qualitative picture (even quantitative sometimes), and this is enough for our discussion here.
In the Debye model, Eq.(\ref{eq:fvib}) reduces to
\begin{equation}
  F_{ph}(V,T)=3k_{B}T\ln\left[1-\exp(-\Theta_{D}/T)\right]-k_{B}Tf\left(\Theta_{D}/T\right)+\frac{9}{8}k_{B}\Theta_{D},
\end{equation}
where $k_B$ is the Boltzmann constant and $f$ is the Debye function. The Debye temperature is
approximated as\cite{geng05,moruzzi88}
\begin{equation}
  \Theta_{D}=\Theta_{D}^{p}\left[\frac{BM^{p}}{B^{p}M}\left(\frac{V}{V^{p}}\right)^{1/3}\right]^{1/2}.
\end{equation}
Here the superscript $p$ denotes the reference state (here the defect-free UO$_{2}$), $B$ is the bulk modulus, $V$ is the effective volume per atom, and $M$ is the effective atomic weight.
This approximation works well for a lot of materials,\cite{geng05,moruzzi88} and is a physically
sound extrapolation for $\Theta_{D}$ of UO$_{2\pm x}$ from that of perfect UO$_{2}$.
The value of the latter, \emph{i.e.} $\Theta_{D}^{p}$, takes 383\,K at ambient conditions according to recent x-ray diffraction analysis.\cite{serizawa00,serizawa99}

The Gibbs free energy for a specific defective structure $i$ is then given by
\begin{equation}
  G_{i}(P,T)=E_{c}(V)+F_{ph}(V,T)+PV.
\end{equation}
For intrinsic point defects, the formation Gibbs free energy (FGE) of a Frenkel pair (FP)
of species $X$ is expressed as
\begin{equation}
  G_{\mathrm{X{\_}FP}}=G^{N-1}_{X_{v}}+G^{N+1}_{X_{i}}-2G^{N},
\end{equation}
and for the Schottky defect ({S}) as
\begin{equation}
  G_{\mathrm{S}}=G^{N-1}_{\mathrm{U}_{v}}+2G^{N-1}_{\mathrm{O}_{v}}-3\frac{N-1}{N}G^{N}.
\end{equation}
If oxygen clusters are considered, we have to introduce isolated oxygen vacancy ($\mathrm{O}_{v}$)
to maintain the composition equation when in the closed regime where no particle exchange
with the exterior occurs. For cuboctahedral clusters (COT), we thus should consider the $\mathrm{O}_{v}$-COT complex
and its FGE reads
\begin{equation}
  G_{\mathrm{COT}}=G^{N+y}_{COT}+yG^{N-1}_{\mathrm{O}_{v}}-(y+1)G^{N},
  \label{eq:Gcoto}
\end{equation}
where $y$ is the excess number of oxygen and takes $5$ in the COT-$o$ cluster.\cite{geng08,geng08b}
In these equations, $N$ is the number of atoms in a defect-free cell (96 for a $2\times2\times2$ supercell)
and $G^{N}$ is the corresponding Gibbs free energy. $G^{N\pm 1}_{X_{v},{X_{i}}}$
and $G^{N+y}_{COT}$
are the Gibbs free energy of the cell containing the respective defect.

It is worthwhile to mention that the COT cluster was originally proposed by
experimentalists to explain the measured neutron diffraction patterns
of U$_{4}$O$_{9}$ \cite{bevan86,cooper04} and U$_{3}$O$_{7}$.\cite{garrido03,nowicki00}
Historically, it was denoted by M$_{6}$X$_{36}$ (or M$_{6}$X$_{37}$ if an
additional anion occupies the center).\cite{bevan86,bevan86b}
However, it is a little misleading since the actual
defect is composed of only 12 interstitials (forming the cuboctahedral geometry)
and eight oxygen vacancies.
There are two possible configurations for COT:
with its center being occupied by another oxygen (COT-$o$) or being empty (COT-$v$).
It was suggested that COT clusters should also present in UO$_{2+x}$.\cite{murray90, geng08,geng08b}
Its geometry, energetics, and structural stability have been discussed in Ref.[\onlinecite{geng08b}].

\begin{figure}
\includegraphics*[width=2.0 in]{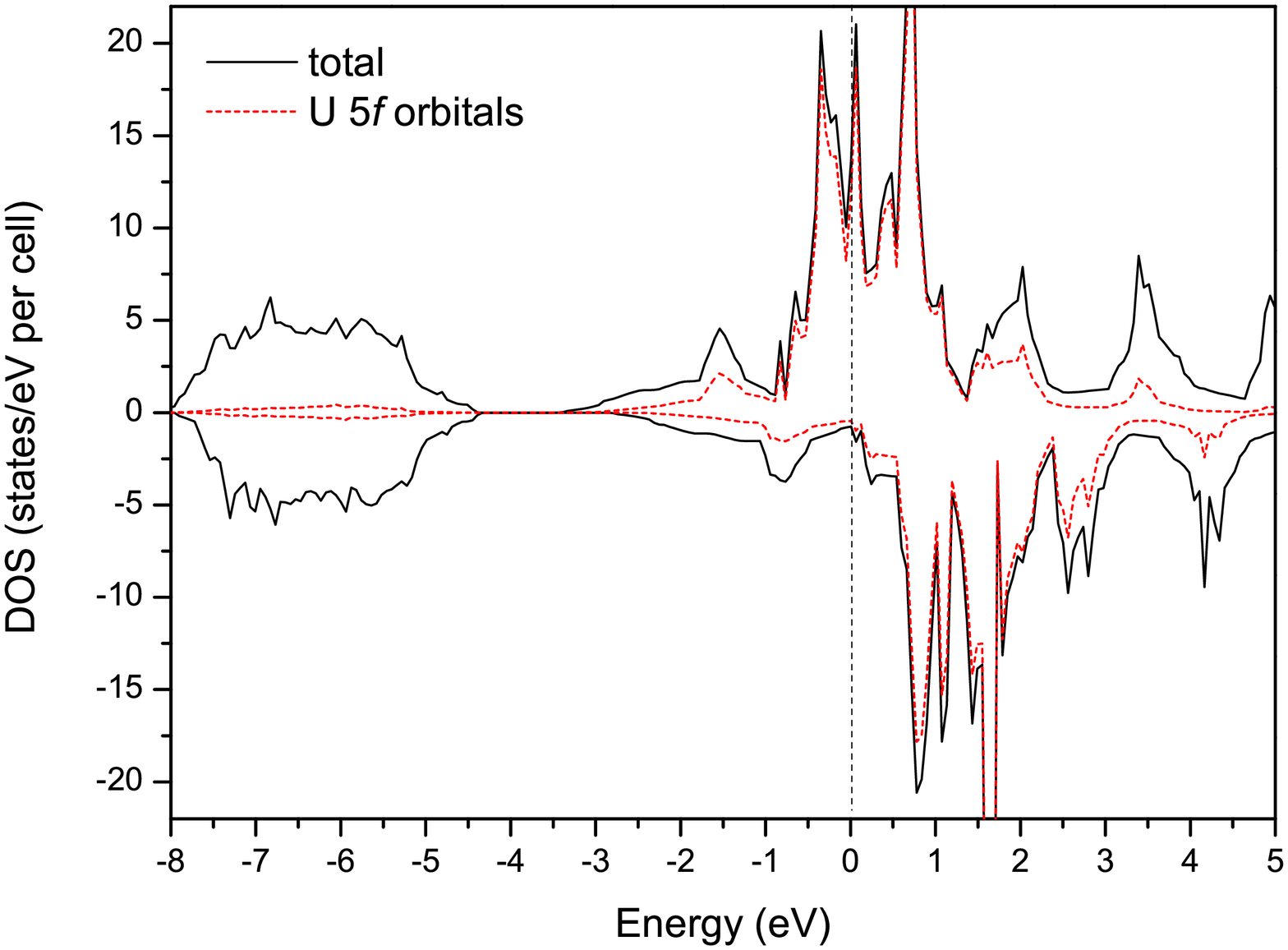}\\
\includegraphics*[width=2.0 in]{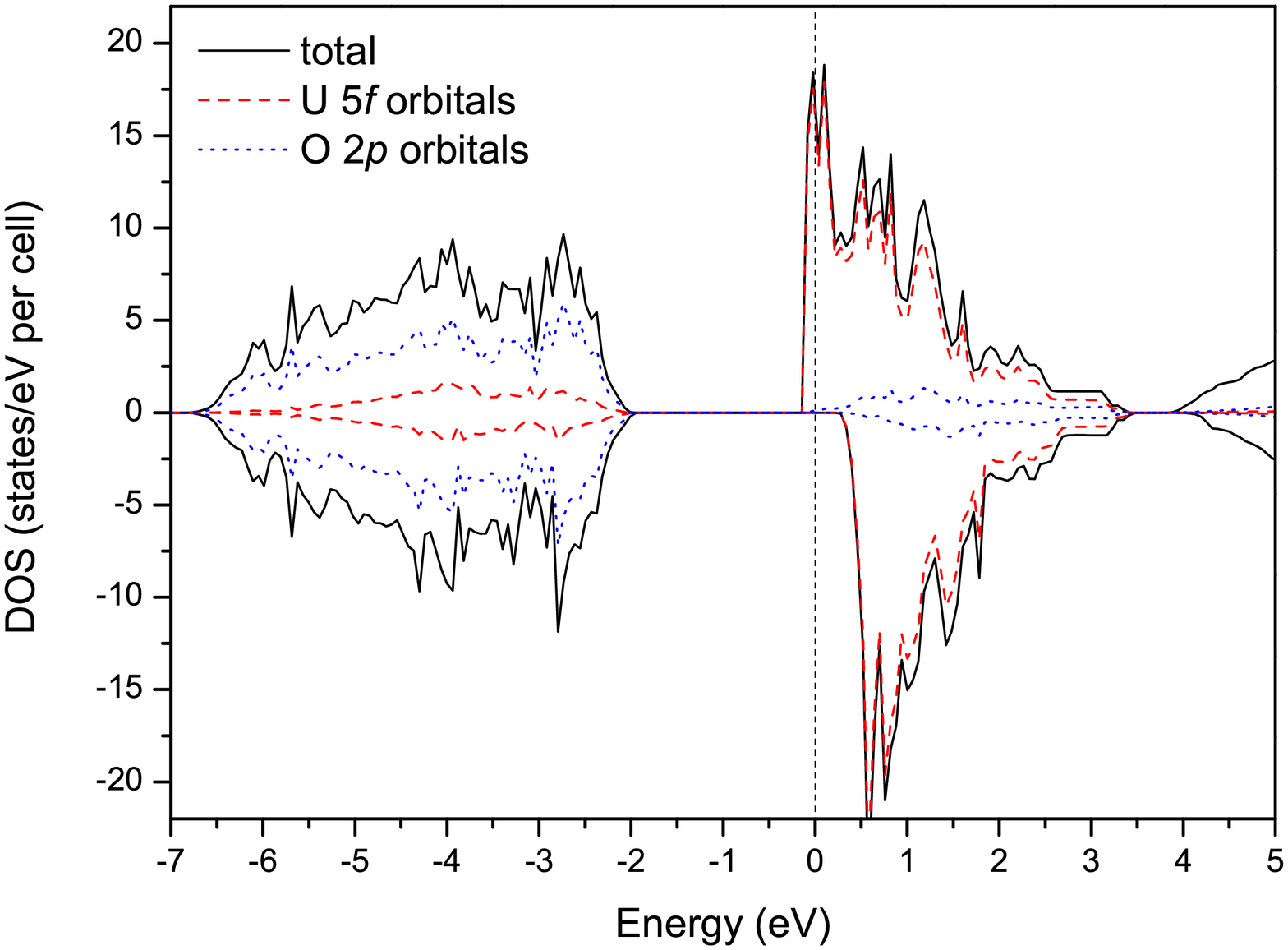}\\
\includegraphics*[width=2.0 in]{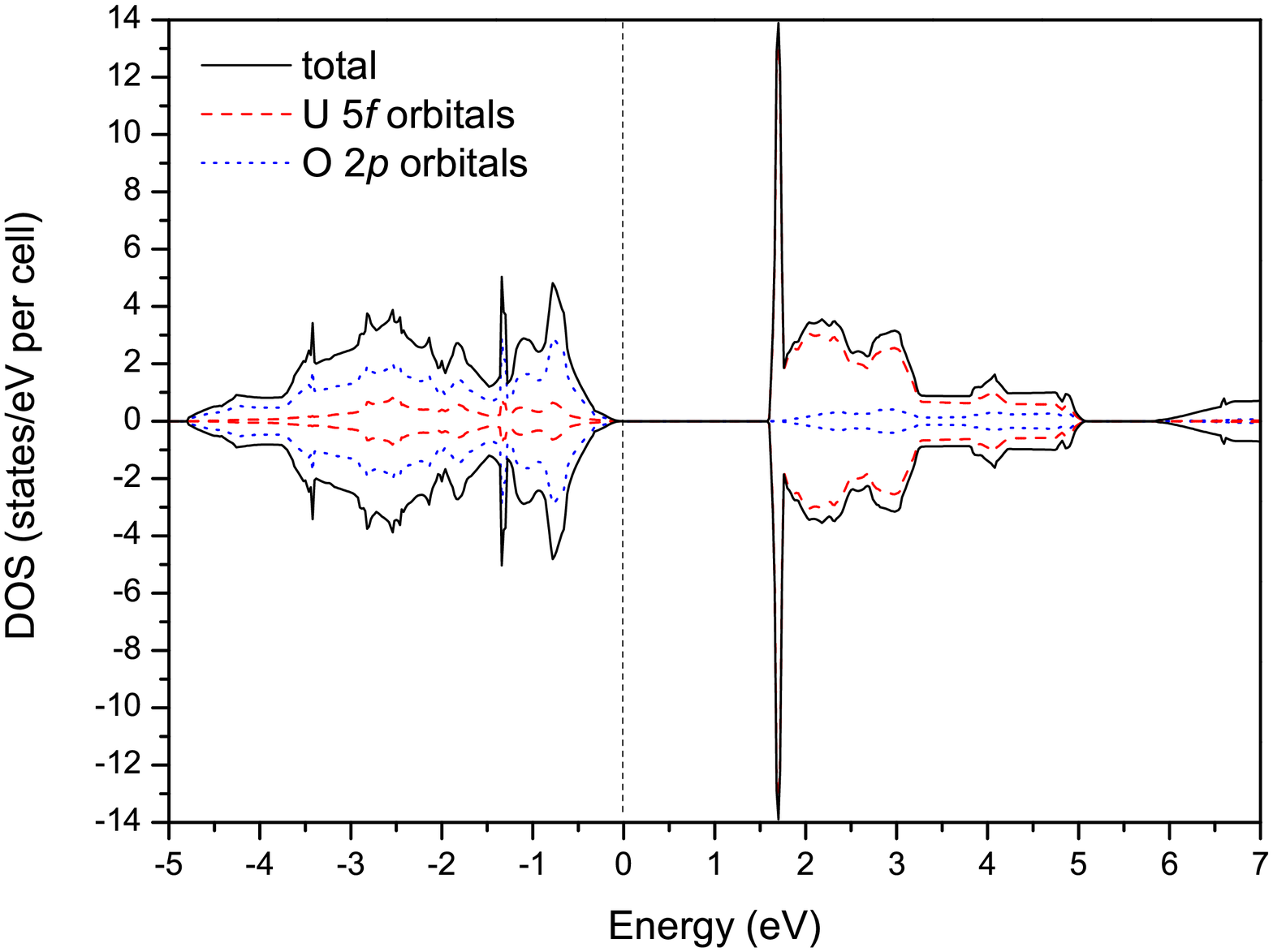}\\
  \caption{(Color online) The total and partial electronic density of states that projected onto the U\,$5f$ and
  O\,$2p$ orbitals
  for UO (upper panel), $\alpha$-U$_{3}$O$_{8}$ (middle panel), and $\delta$-UO$_{3}$
  (lower panel) at 0\,GPa. Vertical dotted lines indicate the position of
  the Fermi levels.}
  \label{fig:UO-dos}
\end{figure}

\section{RESULTS AND DISCUSSION}
\label{sec:EF}
\subsection{Properties at zero Kelvin}
\subsubsection{Uranium monoxide}

In order to understand the general oxidation effects on compression behavior of uranium oxides,
as well as to evaluate the significance of point defects
on EOS of UO$_{2\pm x}$, we studied UO, the $\alpha$ phase of U$_{3}$O$_{8}$, and the $\delta$ phase of UO$_{3}$.
UO is face-centered cubic with the sodium chloride-type structure. The experimental
lattice parameter is $4.92\pm 0.02\,\mathrm{\AA}$.\cite{lam74} However, there have been no claims
of bulk UO having ever been synthesized, and reports of a UO surface phase on U metal for low
exposures to oxygen\cite{ellis76} and UO thin films\cite{lam74} could not be reproduced.\cite{winer86}
Since UO has never
been prepared with great purity, it has been suggested that the observed thin films actually
represent uranium oxynitrides and oxycarbides, which form in the presence of N$_{2}$ or C, and
at low oxygen pressure.\cite{eckle04} The equilibrium lattice parameter and
electronic structure of UO have been studied theoretically by density functional method,\cite{brooks84,petit10}
but no mechanical property and compression behavior were explored, which
are important for the purpose of understanding the response of the materials being oxidized from metal to
dioxide and then to trioxide.

\begin{figure}
  \includegraphics*[width=3.5 in]{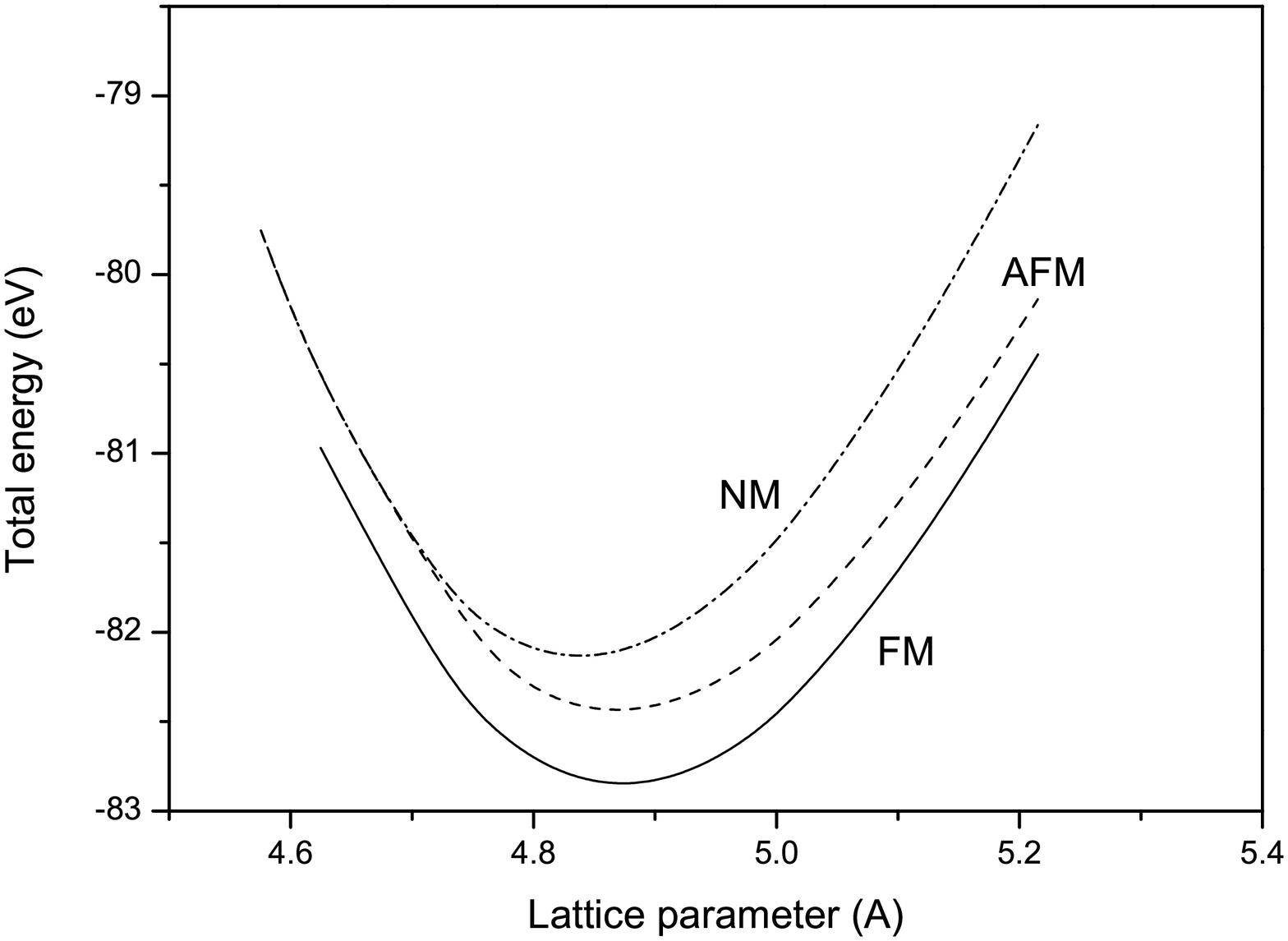}
  \caption{Total energy variation of UO in different magnetic ordering states as a function of the cubic lattice parameter.
  }
  \label{fig:UO-totE}
\end{figure}

Our calculated electronic density of states (DOS) is shown in the upper panel of Fig.\ref{fig:UO-dos}. For UO, the $f$-electron
lingers between the localized and delocalized pictures. Here we assumed itinerant $f$ electrons and
in the ground state it is metallic, as can be seen from the nonzero DOS at the Fermi level. The
O atom has two unoccupied $p$ states, and in the monoxide, the $p$ band can accommodate two electrons
from the uranium atoms ($7s^{2}$) through charge transfer and hybridize with $6p^{6}5f^{3}6d^{1}$ weakly.
The remaining valence electrons,
including $6d$ and partially localized $5f$ electrons, start filling the conduction band, with the Fermi level
pinned to the narrow $f$ band. This agrees with earlier
self-interaction-corrected local spin-density (SIC-LSD) approximation calculations.\cite{petit10}
The energy curves shown in Fig.\ref{fig:UO-totE} confirm that UO is in ferromagnetic (FM) ordering, with
the anti-ferromagnetic (AFM) and non-magnetic (NM) configurations having an energy per cubic cell
about 0.43 and 0.72\,eV higher, respectively. A NM ordering state gives a shorter lattice parameter
($4.84$\,\AA) and a higher bulk modulus (201\,GPa), compared with early LDA results of 4.88\,{\AA}
and 198\,GPa.\cite{brooks84}
Meanwhile AFM and FM ordering have almost the same lattice parameter of 4.88\,\AA, which is in good
agreement with the experimental\cite{lam74} value of 4.92\,{\AA} and
the SIC-LSD\cite{petit10} result of 4.94\,{\AA}.

\subsubsection{Triuranium octoxide and uranium trioxide}

U$_{3}$O$_{8}$ is present as an olive green to black, odorless solid. In the presence
of oxygen, uranium dioxide is oxidized to U$_{3}$O$_{8}$, whereas UO$_{3}$
loses oxygen at temperatures above 773\,K and is reduced to U$_{3}$O$_{8}$.\cite{fujino81}
It is generally
considered to be the more attractive form for disposal purposes of nuclear fuel because, under normal environmental
conditions, U$_{3}$O$_{8}$ is one of the most kinetically and thermodynamically
stable forms of uranium and also because it is the form of uranium found in nature.
At low temperature U$_{3}$O$_{8}$ is in the $\alpha$ phase, a layered orthorhombic structure with
space group $Amm2$,\cite{loopstra70,pillai01} where the layers are bridged by oxygen atoms.
Each layer contains uranium atoms which are in different coordination environments.

\begin{table}
\caption{\label{tab:u3o8} Calculated lattice parameters (in \AA) and cell volume (in \AA$^{3}$) of $\alpha$-U$_{3}$O$_{8}$ compared with
other theoretical and experimental results. }
\begin{ruledtabular}
\begin{tabular}{l l c c c c} %{10cm}{@{\extracolsep{\fill}}l c c c c c}
  Approach && $a$ & $b$ & $c$ & Vol\\
%            &          &  & &  &\\
\hline
PBE &FM, half-metallic& 4.16 & 11.57 & 7.03 &338.36\\
PBE &NM, metallic& 4.16 & 11.77 & 6.87&336.38\\
GGA+\emph{U}\cite{yun11}  &FM, insulator&      4.21 &11.61 &7.20 &351.92\\
Expt.\cite{pillai01}  &Insulator& 4.15 & 11.95 & 6.72&333.26\\
Expt.\cite{loopstra70}&Insulator& 4.14 & 11.96 & 6.72&332.74
\end{tabular}
\end{ruledtabular}
\end{table}

We studied the $\alpha$-U$_{3}$O$_{8}$ phase by computing its atomic geometry,
electronic structure, and energetics.
Table \ref{tab:u3o8} lists the calculated structure parameters by comparing
with GGA+\emph{U}\cite{yun11} and experimental\cite{loopstra70,pillai01} results. It can be seen
that although pure GGA calculations
with PBE functional fail to predict the correct insulator state, the optimized geometry is in
good agreement with experimental measurements, except a slightly shorter $b$ axis and a longer
$c$ axis were obtained. Deviation of the predicted equilibrium volume
from the measured ones is less than 2\%. In contrast, GGA+\emph{U} predicts a much worse result,
with the equilibrium volume about 6\% larger.\cite{yun11} This is quite similar to the embarrassing situation
encountered by the GGA+\emph{U} approach in UO$_{2}$, where a remarkable underestimation of the structural
binding was also obtained.\cite{geng07} Considering the acceptable performance of pure GGA on the energetics
and EOS of UO$_{2}$, we can expect that it will give reasonable results for the properties
of U$_{3}$O$_{8}$
of interest here, even if a wrong half-metallic groundstate was predicted.

The electronic structure of non-magnetic $\alpha$-U$_{3}$O$_{8}$ shows that it is metallic with the Fermi level pinned at the bottom of the conduction band.
The energy gap below it is about 2.2\,eV. The FM ordering phase
has an energy about 0.3\,eV lower per conventional cell than the NM ordering one, and
is half-metallic.
This energy gain is from pushing the minor spin component of the 5$f$ band away from the Fermi level.
However, the remaining 5$f$ electrons on uranium atoms occupy the major spin band
and partially localize at the Fermi level (see the middle panel of Fig.\ref{fig:UO-dos}).
If Hubbard on-site correction were applied,
the narrow 5$f$ band would be completely pushed away from the Fermi level and result in an insulator solution.

Local magnetic moment is a helpful indicator for the localization degree of 5$f$ electrons. It is about $2\,\mu_{B}$
per uranium atom for UO and $1.93\,\mu_{B}$ for UO$_{2}$,\cite{geng07} suggesting there are almost two 5$f$ electrons being partially
localized. The difference is that in UO the localized electrons are at the Fermi level in the conduction band,
whereas in UO$_{2}$ the on-site Hubbard interaction splits the 5$f$ band and leads to
an energy gap (\emph{i.e.}, being Mott insulator), thus the electrons are
localized in the valence band. On the other hand, U$_{3}$O$_{8}$ is close to being a charge-transfer
insulator, in which
5$f$ electrons are absorbed by oxygen to fill the empty $2p$ orbitals. But this process
is incomplete and the remaining partially localized
5$f$ electrons in the conduction band pinned at the Fermi level contribute about $0.7\,\mu_{B}$ per uranium atom,
and make U$_{3}$O$_{8}$ a half-metallic ferromagnet in GGA approximation.
To open up the energy gap, the on-site Coulomb repulsion correction must be involved, and thus
U$_{3}$O$_{8}$ is an intermediate phase between the Mott insulator and the charge transfer insulator,
as shown by the GGA+\emph{U} calculation.\cite{yun11}
This picture is in accordance with nominal chemical valence counting: on average
each uranium atom bonds with $0.67$ more oxygen in U$_{3}$O$_{8}$ than in UO$_{2}$, thus it loses
about $1.3$ more electrons in the former compound. Namely, most parts of the two localized
5$f$ electrons has been transferred to oxygen and only a small portion ($0.7$ electrons) is left behind.
This charge transfer naturally creates a sharp energy gap just below the Fermi level,
resulting in an almost closed shell on the uranium cores. We therefore predict
that UO$_{3}$ would be a pure charge-transfer insulator with the bare uranium ions in the closed-shell Rn configuration.

Calculations on the $\delta$ phase of UO$_{3}$ confirms this argument.
$\delta$-UO$_{3}$ is a cubic phase in ReO$_{3}$ structure with a space group $Pm\overline{3}m$.\cite{weller88}
The lower panel in Fig.\ref{fig:UO-dos} illustrates its calculated electronic DOS.
It is a non-magnetic insulator. No any local magnetic moment on the uranium atoms can be
observed.
A distinct band gap of 1.6\,eV can be clearly seen, which is due to the transfer of
electrons to oxygen atoms, and leaves behind an empty atomic like 5$f$ band.
As a result, the Fermi level moves to the top of the valence band.
This completes the electronic structure evolution of uranium oxides along the oxidation process,
\emph{i.e.}, from metal ($\alpha$-U and UO, NM/FM) to Mott insulator (UO$_{2}$, AFM),
to intermediate insulator (U$_{3}$O$_8$, FM), and then to pure charge transfer insulator (UO$_{3}$, NM) with a gradual shifting
of the 5$f$ electrons from uranium to oxygen atoms (see Fig.\ref{fig:UO-dos}). The correlation effect among 5$f$ orbitals also changes from
one end to the other: in UO$_{2}$ it is the strongest,
whereas in UO$_{3}$ we expect no strongly correlated effect to present at all. This can be
verified by the correct prediction of the insulator groundstate of $\delta$-UO$_{3}$
with the pure GGA approach, as well as
the excellent agreement of the cubic lattice parameter between
the calculated value of 4.168\,{\AA} and the experimental one of 4.165\,{\AA}.\cite{weller88}

\begin{figure}
  \includegraphics*[width=3.0 in]{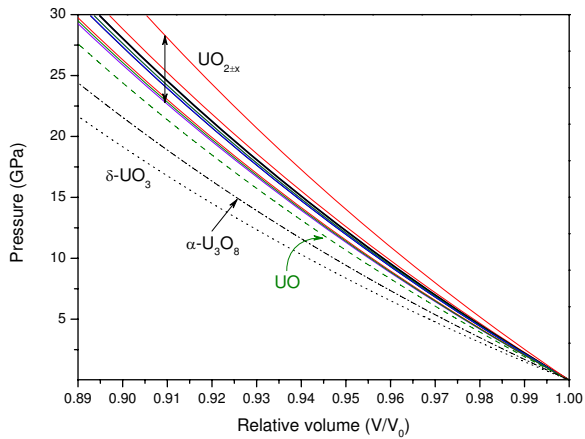}
  \caption{(Color online) Compression curves of UO, $\alpha$-U$_{3}$O$_{8}$, and $\delta$-UO$_{3}$ at zero Kelvin compared with that of UO$_{2\pm x}$;
  the latter spans a quite broad pressure range.
  }
  \label{fig:PV-c}
\end{figure}

\subsubsection{Compressibility of UO$_{2\pm x}$}

The ordered phase of U$_{4}$O$_{9}$ has a complicated structure consisting
of $4\times4\times4$ fluorite-type subcells,\cite{cooper04} which is inaccessible for direct
\emph{ab initio} modeling because of the huge computational demands. We thus
turn to the study of the $C1_{1}$ structure, as well as other defective UO$_{2\pm x}$ configurations as listed in Table \ref{tab:coheE},
and hope they can provide useful information
on the general trend of oxidation effects on compression behavior.
In $C1_{1}$ configuration there is one oxygen interstitial occupying
the largest octahedral hole $(\frac{1}{2},\frac{1}{2},\frac{1}{2})$, and has the same chemical composition
as U$_{4}$O$_{9}$. But one should keep in mind that it is more a defective UO$_{2}$ rather than being U$_{4}$O$_{9}$,
and would overestimate the bulk modulus slightly because of its high symmetry.

The 0\,K compression curves of defective UO$_{2\pm x}$ are shown in Fig.\ref{fig:PV-c}, together
with that of UO, $\alpha$-U$_{3}$O$_{8}$, and $\delta$-UO$_{3}$.
The corresponding parameters for these curves are
listed in Table \ref{tab:coheE}.
It can be seen that although there might be some structure-dependence,
a general trend is evident.
With an increase of the oxidation degree from UO to UO$_{2}$, the compound becomes harder and harder
to compress, with the $C1_{1}$ configuration of UO$_{2\pm x}$ being the stiffest one.
Further oxidization softens the material, and $\delta$-UO$_{3}$ is
the most compressible one.
Please note that $\alpha$-U$_{3}$O$_{8}$ and $\delta$-UO$_{3}$ transform into other structures under
compression at a pressure scale of several GPa.
This makes the $P$-$V$ lines kink at the transition points, and increase the compressibility further.
A detailed analysis of these pressure-induced phase transitions are, however, beyond the scope
of this paper.

\begin{figure}
  \includegraphics*[width=3.5 in]{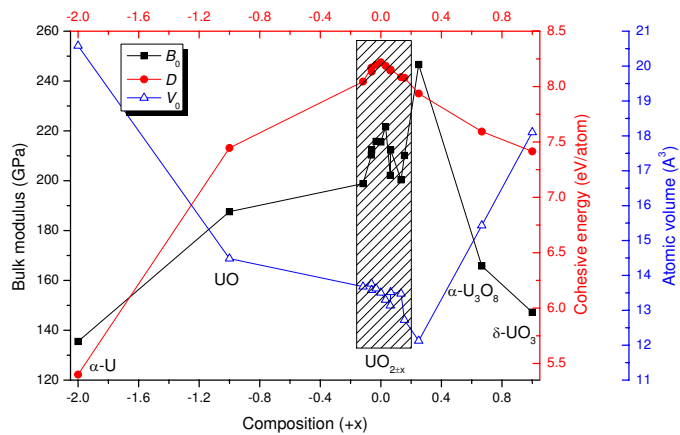}
  \caption{(Color online) Variations of the cohesive energy per atom ($D$), the bulk modulus ($B_{0}$),
  and the atomic volume ($V_{0}$) as a function of composition deviation $x$ at 0\,GPa.}
  \label{fig:BEV}
\end{figure}

The tendency due to oxidation also clearly manifests in Fig.\ref{fig:BEV},
where the correlation between cohesive energy, bulk modulus, and atomic volume is
striking. Note that the values of $\alpha$-U are taken from experiments,\cite{barett63,yoo98} and others are the calculated
results as listed in Table \ref{tab:coheE}.
We can see that stoichiometric UO$_{2}$ has the largest cohesive energy per atom, thus being the most energetically
stable phase, and reflects the fact that when heated to 1573\,K U$_{3}$O$_{8}$ decomposes to UO$_{2}$.
As mentioned above, UO$_{2}$, as well as UO$_{2\pm x}$, lies in the intermediate stage of
the charge transfer process. Therefore it has the strongest covalent bonds that arise from hybridization
between U\,$5f$ and O\,$2p$ orbitals, which then give rise to the smallest atomic volume, the deepest
cohesive energy, and the highest bulk modulus, and therefore the least compressibility of UO$_{2\pm x}$ as shown
in Fig.\ref{fig:PV-c}. Further oxidization to U$_{3}$O$_{8}$ shifts the material quickly into ionic character,
and the covalent bonds become weak. This leads to a huge volume expansion of more than 30\%, and is dangerous
for most nuclear application purposes.

It is worthwhile to point out that in Fig.\ref{fig:PV-c} the $P$-$V$ lines of UO$_{2\pm x}$ span
a wide range of pressure, and is comparable to the difference between stoichiometric UO$_{2}$ and other oxides.
This underlines the importance of an accurate non-stoichiometric EOS,
and it is inappropriate to \emph{represent} it by any specific
defective configuration, especially when the defect species change with temperature and pressure.
Therefore a full EOS model is required for understanding the general thermophysical
properties of off-stoichiometric compounds. That is, one has to study how the stoichiometry variation is
accommodated in the material, what kind of microscopic entities are they in physics, and how these
entities respond to different physical conditions. In the case of UO$_{2}$, it is well known that point defects
as well as some oxygen clusters account for the stoichiometry variation while keeping
the fluorite-type matrix almost unchanged. These features make the construction of the non-stoichiometric
EOS much easier.

\subsection{Finite pressures and temperatures}

Almost all of the earlier defect population analyses were carried out with 0\,K formation
energies.\cite{crocombette01,crocombette11,freyss05,geng08,geng08b,geng08c} It is widely assumed that
the contribution from lattice vibrations is insignificant, but lacks
detailed assessment. Also local residual stress and thermal expansion effects were completely neglected.
The following discussion is devoted to this issue, and reveals that thermal vibrations
have a limited but important influence
on defect populations, whereas the pressure effect is much more striking in comparison.

It is worthwhile to note that since UO$_{2}$ is a semiconductor, the defects can be charged.\cite{nerikar09a,crocombette11}
These charged species might have several charge states.\cite{crocombette11}
These facts and the slow convergence of
a charged system in computation
complicate the discussion (the energy converges linearly with respect
to the inverse of the supercell size
at the first order correction).
Since our main purpose here is to investigate the phonon contribution
and the compression effects on defects behavior, which should be similar for neutral and
charged defects (we are reminded of the qualitative consistency of these two kinds of
defects on 0\,K formation energy\cite{nerikar09a,crocombette11}), we thus focus
our discussion here on neutral defects only. Extension of the analysis
to charged defects is straightforward as long as the energetics of these species are known.

\subsubsection{Variation of formation Gibbs free energy (FGE)}

\begin{figure}
  \includegraphics*[width=3.0 in]{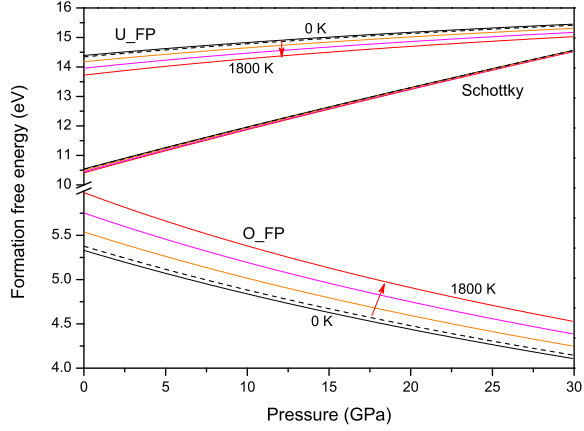}
  \caption{(Color online) Variation of the formation Gibbs free energy of Frenkel pairs and Schottky defect in UO$_{2}$
  under compression at different temperatures. Arrows indicate the direction of temperature increase with
  a step size of 600\,K. Dashed lines correspond to 0\,K results with zero point energy included.
  }
  \label{fig:GF-pd}
\end{figure}

\begin{figure}
  \includegraphics*[width=3.0 in]{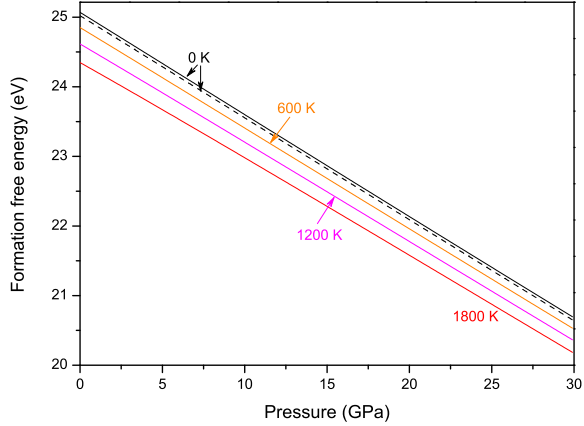}
  \caption{(Color online) Formation Gibbs free energy of the COT-$o$ complex as a function of temperature and pressure.
  }
  \label{fig:GF-coto}
\end{figure}

Variations of FGE as a function of temperature and pressure are
shown in Fig.\ref{fig:GF-pd} for intrinsic point defects (oxygen and uranium Frenkel pairs and Schottky defect, respectively)
and in Fig.\ref{fig:GF-coto}
for the $\mathrm{O}_{v}$-COT-$o$ complex, as defined in Eq.(\ref{eq:Gcoto}).
It can be seen that phonon contribution reduces the FGE of the uranium Frenkel pair (FP), whereas
it enhances that of the oxygen FP.
As a result, its contribution to Schottky defect almost completely disappears
due to cancellation of the opposite effect on uranium and oxygen defects.
The maximum difference of the FGE between 0 and 1800\,K is less than 0.62\,eV for Frenkel
pairs and 0.12\,eV for Schottky defect, respectively. In other words, it counts only about 11\% for oxygen FP, 4\%
for uranium FP, and 1\% for Schottky defect.

The calculated FGE at 0\,GPa and 0\,K are consistent with that of Ref.[\onlinecite{geng08}].
But we have to point out that there is a typo in Ref.[\onlinecite{geng08}] where
the formation energy of the uranium point interstitial (U-Int.) should be 5.642\,eV
instead of 8.194\,eV. This change affects Table I and Table IV in Ref.[\onlinecite{geng08}].
Also, it reduces the energy of the uranium Frenkel pair (U-FP) in Table IV from 17.2
to 14.6\,eV. However, this does not affect the defect population analysis since
the concentration of uranium interstitial is too small to have any significant contribution
to the stoichiometry of UO$_{2}$.

On the other hand, lattice vibrations decrease the FGE
of the COT-$o$ complex (see Fig.\ref{fig:GF-coto}). Increasing the temperature from 0 to 1800\,K, the change in the FGE of the COT-$o$ complex is less than 0.72\,eV,
which is about 3\% of the total formation energy. Compression depresses the
phonon effects slightly. But totally, the thermal vibrational contribution to the FGE of point defects
and the oxygen defect cluster is small.

In contrast, the compression effect is quite prominent. Within a pressure range of 30\,GPa, which
is easy to achieve with diamond-anvil cell (DAC) devices or micro shock-wave loadings in nuclear
materials driven by thermal spikes or collision cascade events arisen from fission reactions,
the FGE increases 1.07, $-1.23$, and 4.02\,eV for uranium FP, oxygen FP, and Schottky defects,
and amounts to 7\%, $-23$\%, and 38\% of the zero pressure values, respectively.
For the COT-$o$ complex, the FGE change is $-4.37$\,eV, about $-17$\% of the 0\,GPa value.
Note that a pressure of
30\,GPa corresponds to an energy scale of 2.4\,eV per atom in UO$_{2}$.

\subsubsection{Defect populations}

\begin{figure}
  \includegraphics*[width=3.0 in]{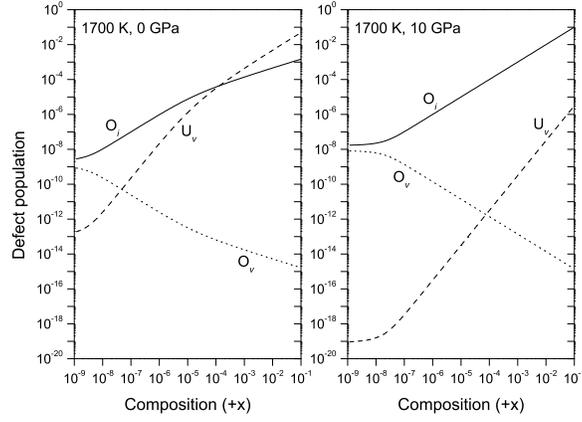}
  \caption{Variation of concentrations of point defects at a temperature of 1700\,K with $x$,
  the deviation from the stoichiometry of UO$_{2}$ in the hyper-stoichiometric regime: left is at a pressure of 0\,GPa
  and right is under 10\,GPa. Solid (dotted and dashed) line indicates the
  concentration in oxygen interstitial (oxygen vacancy and uranium vacancy). The concentration
  of uranium interstitial is negligible.}
  \label{fig:pdm-conc}
\end{figure}

In addition to being a key content in the construction of a non-stoichiometric EOS,\cite{yakub06}
the importance of analyzing defect populations, or equivalently the concentrations, also lies in the challenge of
the elusive physics of UO$_{2+x}$.
Although experiments have shown that COT-$o$ clusters constitute the ordered phase of U$_{4}$O$_{9}$
on a fluorite matrix,\cite{hering52,bevan86,cooper04} it is still in debate whether or not such a cluster also appears
in UO$_{2+x}$.\cite{willis64b,willis78,murray90,geng08b,geng08c} Calculations using 0\,K formation energy suggested that the existence of COT-$o$
in UO$_{2+x}$ is quite likely, but the observed ratio of oxygen interstitials occupying different sites cannot be accounted
for quantitatively.\cite{geng08b,geng08c} This implies that we cannot exclude other possibilities.
In this sense it is very important to figure out how the population of each possible defect changes with
temperature and pressure.
On the other hand, as mentioned above, one of the GGA+\emph{U} calculations predicted a positive 0\,K formation energy
for point oxygen interstitial.\cite{dorado10} If this reflects the true physical behavior, we can expect
that phonon contribution at elevated temperatures should be as such that it makes oxygen defects
the major component. These considerations underline the necessity to investigate
the defect populations at different pressures and temperatures with phonon contributions having been taken into
account.

Using the point defect model (PDM) developed by Lidiard and Matzke,\cite{lidiard66,matzke87,geng08,geng08b}
defect populations around the dilute limit
can be computed.
The key point is that here
the FGE instead of the 0\,K formation energies are employed.
In the following we mainly discuss the hyperstoichiometric ($x>0$) regime because
in the other side of $x<0$, it is always oxygen vacancy (O$_{v}$) that predominates.
Although both pressure and temperature enhance the concentration of point oxygen interstitial (O$_{i}$) and
uranium interstitial (U$_{i}$) slightly, they are still at least two orders smaller
than that of O$_{v}$ and can be neglected safely.

Figure \ref{fig:pdm-conc} illustrates the defect populations as a function of composition at 1700\,K
under pressures of 0 and 10\,GPa, respectively. By comparing with Fig.11 in Ref.[\onlinecite{geng08}]
where only 0\,K and 0\,GPa formation energies were used in the PDM analysis, we find a slight decrease of
the concentration of O$_{i}$ and an increase of the uranium
vacancy (U$_{v}$), and the latter becomes the major component when $x>10^{-3}$. Both
lattice vibrations and thermal expansion contribute to this change.
From Fig.\ref{fig:GF-pd} we learn that
phonon favors uranium defects. Thermal expansion also
prefers uranium FP and Schottky defect because at
an elevated temperature thermal expansion leads to a larger volume that corresponds to a negative cold pressure, which
reduces the formation energy of uranium FP and Schottky but lifts that of oxygen FP.
This result is strongly against the experimental observation that oxygen defects are predominant
in UO$_{2}$ at any temperature and stoichiometry.\cite{matzke87,willis64b,willis78,murray90}
Increasing pressure can depress the uranium vacancy, as shown in the right panel of Fig.\ref{fig:pdm-conc}.
This is easy to understand by inspecting the variation of FGE with pressure
as shown in Fig.\ref{fig:GF-pd}. Nevertheless, this compression effect is not enough to explain the
observed predominance of oxygen defects because the possible residual stress within the fuel
is usually less than 1\,GPa, and at such a pressure scale the uranium vacancy is still comparable
to oxygen defects.

\begin{figure}
  \includegraphics*[width=3.0 in]{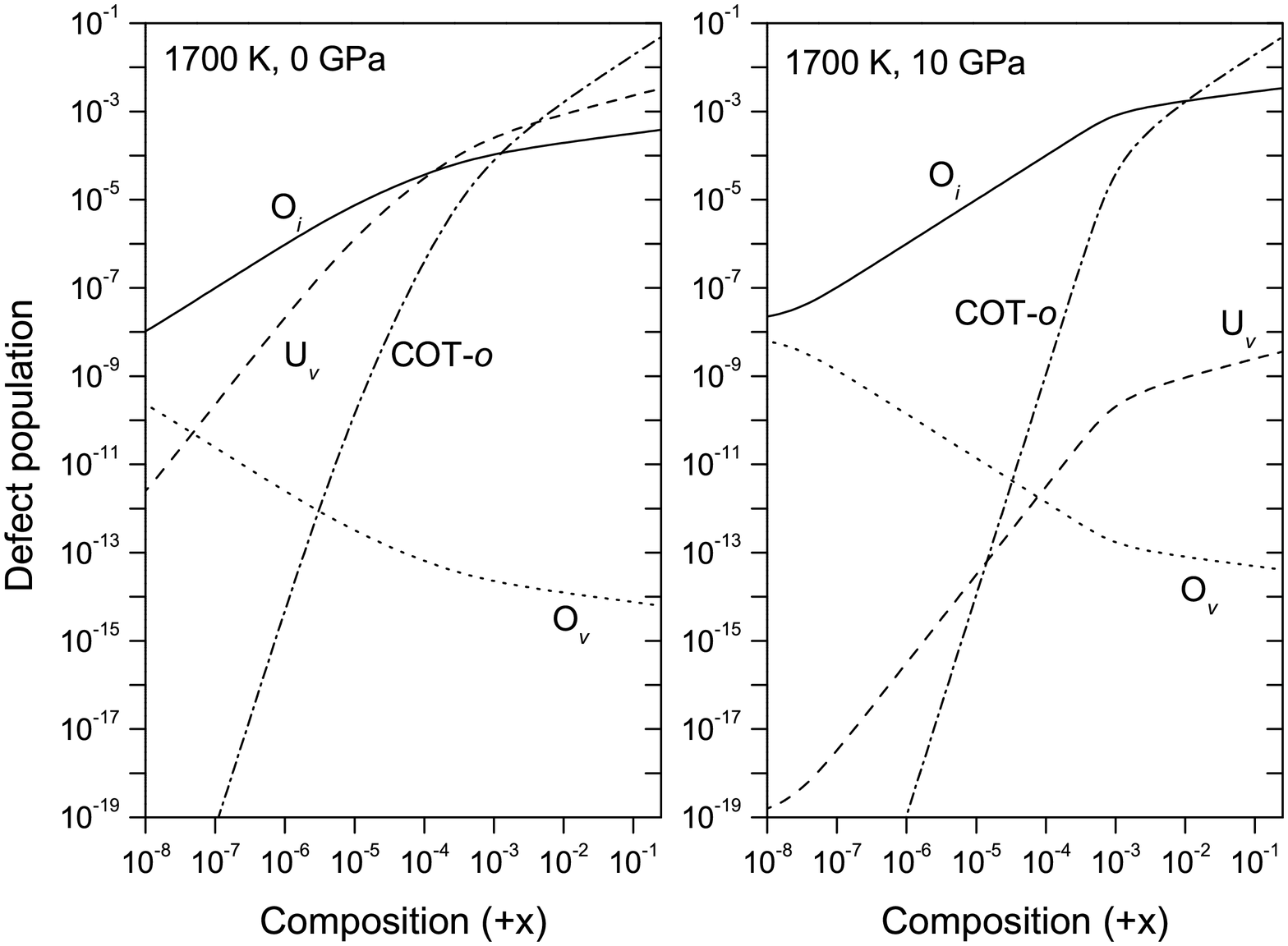}
  \caption{The influence of oxygen cluster COT-$o$ on defect concentrations at 0 and 10\,GPa,
  respectively. The dash-dotted line indicates the concentration of COT-$o$ cluster;
  others are the same as in Fig.\ref{fig:pdm-conc}.}
  \label{fig:ICA}
\end{figure}

\begin{figure}
  \includegraphics*[width=3.0 in]{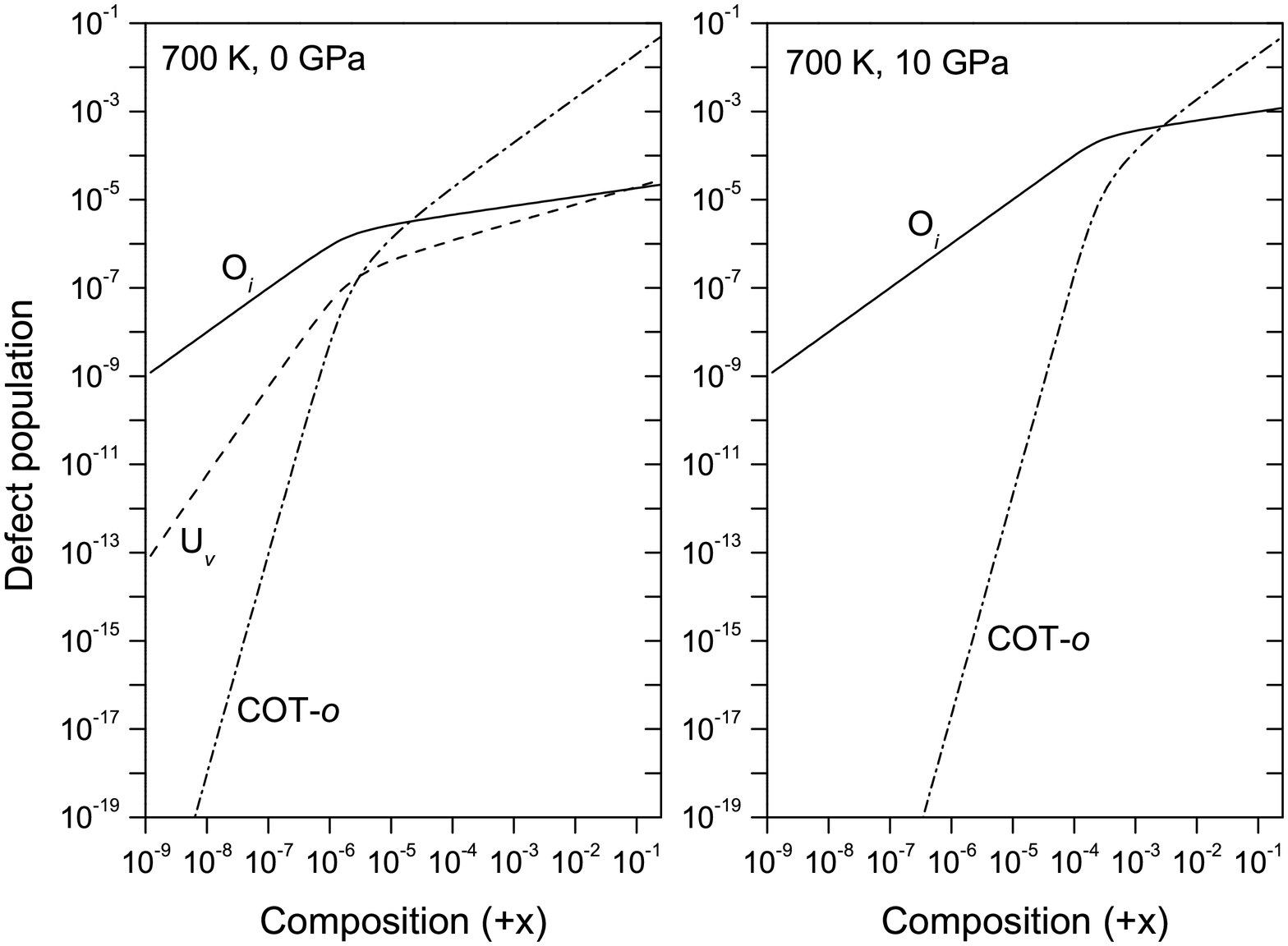}
  \caption{The same as in Fig.\ref{fig:ICA} but now with a temperature of 700\,K.}
  \label{fig:ica-700k}
\end{figure}

It thus becomes interesting to investigate whether phonon and compression effects can alter the
behavior of the COT-$o$ cluster greatly or not. Earlier studies showed that when only 0\,K formation energies were
used, the COT-$o$ cluster is dominant at low temperature and high composition region.\cite{geng08b,geng08c}
Figure \ref{fig:ICA} plots the variation of defect populations at 1700\,K with COT-$o$ included.
Note here that we didnot consider the internal degeneracy of this cluster,\cite{geng08b} since physically no two
clusters can occupy the same lattice site simultaneously if we ignore quantum effects and treat oxygen
as a classical particle.
We can see that, as in the previous case, phonon contribution enhances uranium vacancy, which
is very clear if comparing with Fig.3 of Ref.[\onlinecite{geng08b}].
The earlier conclusion that COT-$o$ is predominant is still preserved when $x>10^{-2}$.
At a pressure of 10\,GPa,
uranium vacancy is drastically suppressed, leaving only single oxygen interstitial and COT-$o$
in competition (the right panel of Fig.\ref{fig:ICA}).
Lowering the temperature to 700\,K further suppresses uranium vacancy and oxygen interstitial, as Fig.\ref{fig:ica-700k}
shows, meanwhile COT-$o$ gets some promotion.
Increasing the pressure to 10\,GPa then sweeps all other defects away and leaves only O$_{i}$ and
COT-$o$ clusters, with a slight depression of the latter and an enhancement of the former.

It is evident that lattice vibrations and hydrostatic compression change the competition among defects greatly,
and the conclusions obtained with 0\,K and 0\,GPa formation energies need be revised:
when considering only point defects, the
earlier predicted predominance of O$_{i}$ doesnot hold anymore with the full FGE,
and U$_{v}$ becomes the dominant defect when $x>10^{-3}$.
That is to say, at the full level
of the theory the experimental fact cannot be reproduced if only
neutral point defects are considered.
On the other hand, when the COT-$o$ cluster is taken into account,
the predominance of this oxygen cluster can be easily established both
at the full FGE level and at the 0\,K and 0\,GPa formation energy level.
The calculated results clearly demonstrate
that hydrostatic pressure has a strong influence on defect populations, and oxygen clusters
are definitely necessary in order to suppress the concentration of uranium vacancy to match the experimental observation qualitatively.

Considering the low temperature phase diagram\cite{higgs07} of UO$_{2+x}$ and the negative formation free energy
of O$_{i}$ and COT-$o$, it becomes clear that the possibility of phase separation between UO$_{2}$ and
U$_{4}$O$_{9}$ must be considered seriously if one wants to understand the underlying physics correctly.
Because U$_{4}$O$_{9}$ can be viewed as an ordered phase of COT-$o$ clusters on a fluorite
matrix,\cite{bevan86,cooper04} this phase separation is in fact a process of formation and segregation of
oxygen clusters. Thus at the intermediate composition and elevated temperature range where
UO$_{2+x}$ is stable, the dynamics of decomposition/formation of COT-$o$ is crucial,
as well as other competitive clusters such as ($V$-3O$^{''}$)$_{2}$.\cite{andersson09,andersson09b,geng08c,geng2010}
It is worthwhile to highlight the connection between $V$-3O$^{''}$ based
clusters\cite{andersson09,andersson09b,geng08c} and cuboctahedral clusters\cite{bevan86,cooper04,murray90,geng08b} (mainly COT-$o$).
Although the latter is more
stable at low temperature,\cite{geng08b,geng08c} it was predicted that with
increasing temperature COT-$o$ will gradually dissociate into a set of single point defects.\cite{geng08b}
The intermediate state of this process is of great importance for understanding the kinetic
behavior. The relatively low energy of ($V$-3O$^{''}$)$_{n}$ makes it
a promising candidate of this transition state.\cite{geng2010} This is possible
because the COT cluster in fact can be viewed as a complex formed by eight tightly connected
$V$-3O$^{'}$ clusters (a distorted version of $V$-3O$^{''}$). Then partial dissociation
of COT clusters will likely go through a configuration of $V$-3O$^{''}$/O$^{'}$
kinetically to lower the energy barrier; that is,
$$\text{COT-}o\Longleftrightarrow (V\text{-3O}^{''})_{n}\Longleftrightarrow 5\text{O}_{i}.$$

\subsubsection{Non-stoichiometric equation of state and shock Hugoniot}

Off-stoichiometry is widely observed in materials of many categories,
for example, in ionic compounds (\emph{e.g.}, oxides, nitrides, and carbides), alloys and intermetallic compounds,
doped semiconductors, and high-temperature superconductors, and so on.
In these materials, the underlying structure of the matrix is usually unchanged or just slightly distorted,
but the electronic
and magnetic properties can be controlled and tuned for application purposes.
The maximal stoichiometry deviation indicates the capability
of the matrix structure to tolerate distortions caused by
doped elements or defects. The mechanical properties usually are determined by the matrix,
but are also varied with respect to different chemical compositions.

The thermodynamic behavior of these materials across a wide range of temperature and pressure
are of interest not only for fundamental physics and material science but also for engineering applications,
in which EOS is one of the most important contents.
There is an elegant theory developed on Ising models of lattice gases, namely, the cluster variation
method (CVM),\cite{kikuchi51} which describes the phase diagram and EOS of binary and pseudo-binary alloys
very well.\cite{connolly83,geng05,geng05b,sluiter90,sluiter96} To generalize this method beyond the ternary system is difficult.
For ionic compounds, however, the stoichiometry usually is accommodated via defects, and is
a multi-component system with the number of species greater than four. Furthermore, not all of these
defects sit on the lattice site of the matrix, which complicates the construction of the
theory, and slows down the convergence even if a CVM-like theory could be constructed.\cite{geng08}
From this point of view, a different route should be attempted for a practical \emph{ab initio} EOS model
(\emph{i.e.}, without any empirical parameters) of non-stoichiometric compounds.

Having known the defect concentrations,
we can express the total Gibbs free energy $G$ of a non-stoichiometric material at pressure
$P$ and temperature $T$ as a function of the defect concentrations $n_{i}$ that accommodate the stoichiometry deviation.
Around the vicinity
of the dilute limit, Taylor expansion of $G$ with respect to $n_{i}$ reads\cite{note01}
\begin{equation}
  G=G^{p} + \sum_{i}A_{i}n_{i}+\frac{1}{2}\sum_{i,j}B_{ij}n_{i}n_{j}+\cdots,
  \label{eq:gibbs}
\end{equation}
where $G^{p}$ is the Gibbs free energy of the matrix, $A_{i}$ is the contribution of single defects,
and $B_{ij}$ is the interaction strength between defects, and so on.
At the first order approximation we can express the non-stoichiometric
Gibbs free energy by making use of specific defective structures as reference;
that is,
\begin{equation}
  G\approx G^{p}+\sum_{i}\frac{\Delta G_{i}}{n_{i}^{\mathrm{ref}}}n_{i}.
  \label{eq:Gtot}
\end{equation}
Here the superscript ``$\mathrm{ref}$'' indicates the
defective reference system, $n_{i}$ is the concentration of
defect type $i$, and the Gibbs free energy difference is $\Delta G_{i}=G_{i}^{\mathrm{ref}}-G^{p}$.
Obviously this linear approximation assumes that the interaction term between
defects $B_{ij}$ is insignificant, which is only true at the dilute limit. However, making use of
reference states extends the applicability of this approximation greatly. It is evident from
Eq.(\ref{eq:Gtot}) that $G$ is close to being exact when $n_{i}=n_{i}^{\mathrm{ref}}$.
Nevertheless, within this linear approximation, the solubility of defects and spinodal decomposition of the system
cannot be correctly described, where the nonlinear terms in Eq.(\ref{eq:gibbs}) are essential.

Within this framework, the system volume becomes
\begin{equation}
  V=V^{p}+\sum_{i}\frac{V_{i}^{\mathrm{ref}}-V^{p}}{n_{i}^{\mathrm{ref}}}n_{i},
\end{equation}
and this gives the equation of state for a non-stoichiometric material. Defect concentrations $n_{i}$
at thermodynamic equilibrium conditions are
determined with the composition equation using the
point defect model as discussed above.

Hugoniot is a set of locus of thermodynamic states that describe the response of a material
to shock wave loadings. It is governed by the law of energy conservation across the
discontinuous zone generated by shock waves, \emph{i.e.},
\begin{equation}
  E-E_{0}=\frac{1}{2}\left(P+P_{0}\right)\left(V_{0}-V\right),
  \label{eq:hugoniot}
\end{equation}
and serves as an important content of high-pressure high-temperature EOS.
For theoretical analysis, it is convenient
to reformulate Eq.(\ref{eq:hugoniot}) as\cite{geng05b}
\begin{equation}
  G-G_{0}=\frac{1}{2}\left(P-P_{0}\right)\left[\left(\frac{\partial G}{\partial P}\right)_{T,\,n_{i}}+
  \left(\frac{\partial G_{0}}{\partial P}\right)_{T,\,n_{i}}
  \right] + T\left(\frac{\partial G}{\partial T}\right)_{P,\,n_{i}}
  -T_{0}\left(\frac{\partial G_{0}}{\partial T}\right)_{P,\,n_{i}},
\end{equation}
where $G_{0}$ is the system Gibbs free energy at an initial condition of pressure $P_{0}$
and temperature $T_{0}$.

\begin{figure}
  \includegraphics*[width=3.0 in]{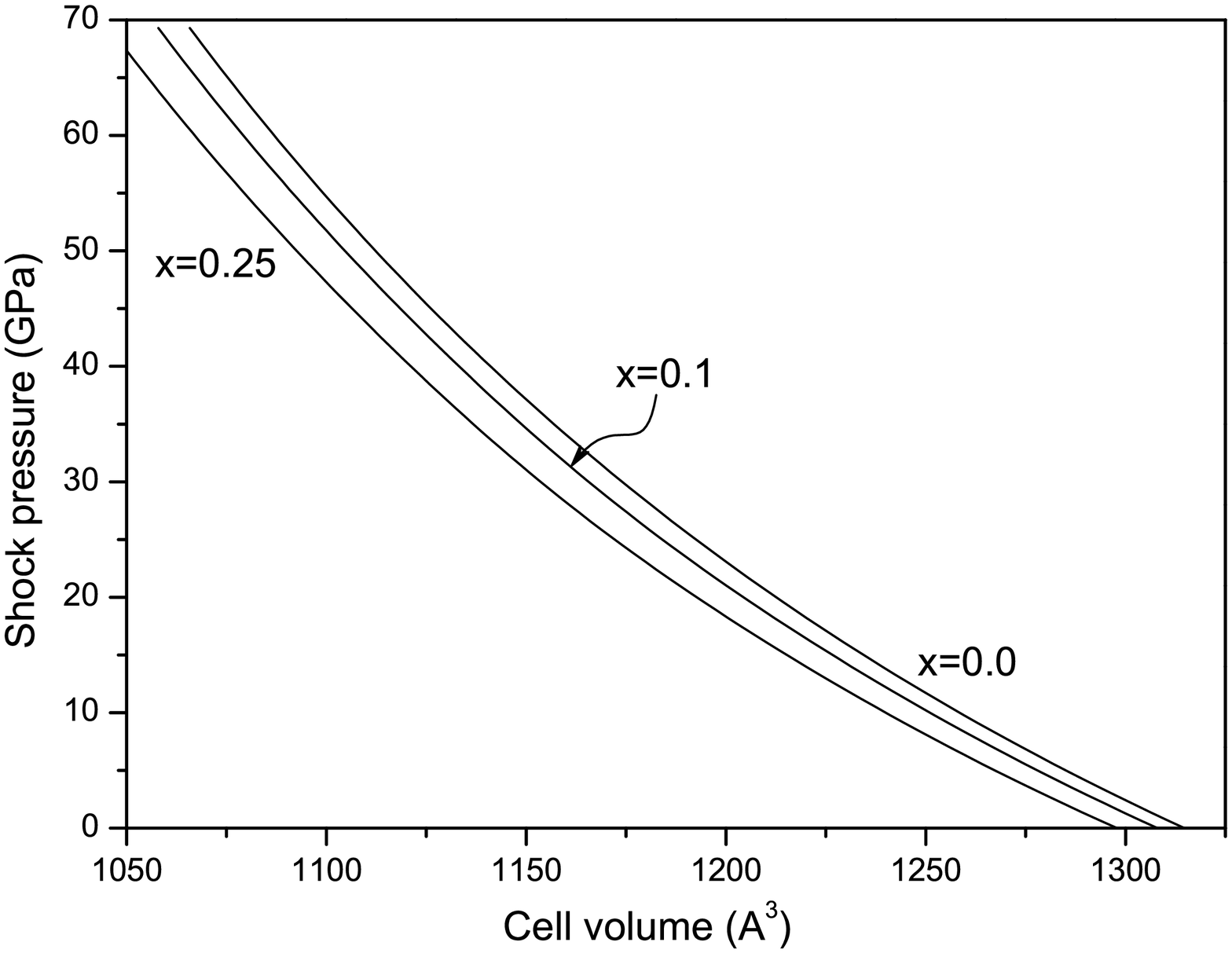}
  \caption{Hugoniot of UO$_{2+x}$ shocked from an initial state of 500\,K and 0\,GPa.
  }
  \label{fig:shock}
\end{figure}

Taking the largest defective configurations as the reference structures (\emph{i.e.}, $C8$ and COT-$o$ configurations
as listed in Table \ref{tab:coheE} that are based on a $2\times2\times2$ supercell of fluorite cubes),
the shock Hugoniot of UO$_{2+x}$ up to 70\,GPa were calculated and shown in Fig.\ref{fig:shock} with
$x=0$, 0.1, and 0.25, respectively.
Note that since here all defects are defined on a fluorite cubic cell of UO$_{2}$, the Gibbs free energy $G$ must be given with
respect to the same cell.
The initial condition of shock is at 0\,GPa and 500\,K.
The highest shock temperature is less than 1200\,K,
reflecting the high bulk modulus of the material and thus a weak energy accumulation of shock waves.
Although the cubic fluorite phase of UO$_{2}$ transforms
into orthorhombic $Pnma$ at room temperature beyond 42\,GPa,\cite{idiri04,geng07} the transition is rather
sluggish and the fluorite phase persists up to 70\,GPa.\cite{idiri04} This suggests the selected range of pressure
in Fig.\ref{fig:shock} is appropriate. Beyond that pressure, we must reanalyze the
geometry of defects and their occupation dynamics because the matrix has already been changed due to phase transition.

From Fig.\ref{fig:shock} we can see that the Hugoniot of UO$_{2+x}$ is very similar to stoichiometric UO$_{2}$,
with the whole curve shifting to the left and reducing the cell volume correspondingly with an increase of $x$.
This is reasonable because from the above analysis we learned that within this pressure and
temperature range the COT-$o$ cluster is always predominant, and the defective structure with
this cluster has a very similar 0\,K energy curve as that of UO$_{2}$ (see Table \ref{tab:coheE}).
However, this is not always the case for non-stoichiometric compounds. Along the shock wave loading path,
pressure and temperature change from point to point because of energy accumulations.
If species of the major defect component were changed due to this kind of thermodynamic condition
variation,
the atomic scale behavior of defects would then manifest itself on the Hugoniot curve.
In that case the non-stoichiometric EOS
would be a useful tool to study this phenomenon theoretically and thoroughly.

\section{CONCLUSION}

We have presented calculations, using the LSDA+\emph{U} exchange-correlation functional,
of the high-temperature high-pressure properties of point defects in UO$_{2\pm x}$.
The phonon contributions were taken into account via the Debye model. The results showed
that lattice vibrations enhance the population of uranium vacancy slightly, whereas
hydrostatic compression
changes the energetics of all defects and then their populations greatly. At ambient pressure, however, inclusion of phonon free
energy and thermal expansion effect fails to reproduce the experimental observation that
oxygen defects should be predominant at the $x>0$ regime, suggesting that the inclusion of defect clusters such as COT-$o$
is necessary.
Furthermore, both lattice vibrations and compression favor the COT-$o$ cluster, and its predominance
in the hyper-stoichiometric regime gets reinforced as compared with the results obtained with 0\,K formation energies.

In order to understand the overall trend of oxidation effects,
the 0\,K electronic structures and equations of state
of UO, UO$_{2\pm x}$, $\alpha$-U$_3$O$_{8}$, and $\delta$-UO$_{3}$ were investigated thoroughly,
revealing the transition of the groundstate of uranium oxides from being metallic to Mott insulator and then to charge-transfer insulator
across the full oxidation degree.
Variations of cohesive energy, atomic volume, and stiffness of the oxides due to evolution of the chemical
bonds from being metallic to covalent and then to ionic characteristics were well demonstrated.
The necessity to develop an EOS model for off-stoichiometric
compounds was also highlighted by comparing the compression curves of defective UO$_{2\pm x}$ with
that of other uranium oxides.
With the knowledge of defect populations within a wide range of
pressure and temperature, a specific equation of state for non-stoichiometric compounds was constructed.
The shock Hugoniots of UO$_{2+x}$ were then studied to illustrate the performance of this model.

\begin{acknowledgments}
Support from the Fund of National Key Laboratory of Shock Wave and Detonation Physics of China
(under Grant No. 9140C6703031004) is acknowledged.
\end{acknowledgments}

%\appendix*
%\section{Structures}
%\end{center}


\begin{thebibliography}{99}
%-------------------
\bibitem{idiri04} M. Idiri, T. Le Bihan, S. Heathman, and J. Rebizant, Phys. Rev. B \textbf{70}, 014113 (2004).
\bibitem{geng07}  H. Y. Geng, Y. Chen, Y. Kaneta, and M. Kinoshita, Phys. Rev. B \textbf{75}, 054111 (2007).
\bibitem{catlow81} C. R. A. Catlow, in \emph{Nonstoichiometric Oxides}, edited by O. T. S{\o}rensen (Academic, New York, 1981).
\bibitem{willis64b} B. T. M. Willis, Proc. Br. Ceram. Soc. \textbf{1}, 9 (1964).
\bibitem{willis78} B. T. M. Willis, Acta Crystallogr., Sect. A: Cryst. Phys., Diffr., Theor. Gen. Crystallogr. \textbf{A34}, 88 (1978).

\bibitem{crocombette01} J. P. Crocombette, F. Jollet, T. N. Le, and T. Petit, Phys. Rev. B \textbf{64}, 104107 (2001).
\bibitem{freyss05} M. Freyss, T. Petit, and J. P. Crocombette, J. Nucl. Mater. \textbf{347}, 44 (2005).
\bibitem{iwasawa06} M. Iwasawa, Y. Chen, Y. Kaneta, T. Ohnuma, H. Y. Geng, and M. Kinoshita, Mater. Trans. \textbf{47}, 2651 (2006).
\bibitem{geng08}  H. Y. Geng, Y. Chen, Y. Kaneta, M. Iwasawa, T. Ohnuma, and M. Kinoshita, Phys. Rev. B \textbf{77}, 104120 (2008).
\bibitem{nerikar09a} P. Nerikar, T. Watanabe, J. S. Tulenko, S. R. Phillpot, and S. B. Sinnott, J. Nucl. Mater. \textbf{384}, 61 (2009).
\bibitem{yu09} J. Yu, R. Devanathan, and W. J. Weber, J. Phys.: Condens. Matter \textbf{21}, 435401 (2009).
\bibitem{andersson09b} D. A. Andersson, T. Watanabe, C. Deo, and B. P. Uberuaga, Phys. Rev. B \textbf{80}, 060101(R) (2009).
\bibitem{gupta10} F. Gupta, A. Pasturel, and G. Brillant, Phys. Rev. B \textbf{81}, 014110 (2010).
\bibitem{dorado10} B. Dorado, G. Jomard, M. Freyss, and M. Bertolus, Phys. Rev. B \textbf{82}, 035114 (2010).
\bibitem{crocombette11} J. P. Crocombette, D. Torumba, and A. Chartier, Phys. Rev. B \textbf{83}, 184107 (2011).

\bibitem{catlow77} C. R. A. Catlow, Proc. R. Soc. London, Ser. A \textbf{353}, 533 (1977).
\bibitem{grimes91} R. W. Grimes and C. R. A. Catlow, Philos. Trans. R. Soc. London, Ser. A \textbf{335}, 609 (1991).
\bibitem{crocombette02} J. P. Crocombette, J. Nucl. Mater. \textbf{305}, 29 (2002).
\bibitem{freyss06} M. Freyss, N. Vergnet, and T. Petit, J. Nucl. Mater. \textbf{352}, 144 (2006).
\bibitem{nerikar09} P. V. Nerikar, X. Y. Liu, B. P. Uberuaga, C. R. Stanek, S. R. Phillpot,
and S. B. Sinnott, J. Phys.: Condens. Matter \textbf{21}, 435602 (2009).
\bibitem{geng2010} H. Y. Geng, Y. Chen, Y. Kaneta, M. Kinoshita, and Q. Wu, Phys. Rev. B \textbf{82}, 094106 (2010).

\bibitem{petit96} T. Petit, B. Morel, C. Lemaignan, A. Pasturel, and B. Bigot, Philos. Mag. B \textbf{73}, 893 (1996).
\bibitem{dudarev97} S. L. Dudarev, D. N. Manh, and A. P. Sutton, Philos. Mag. B \textbf{75}, 613 (1997).
\bibitem{dudarev98} S. L. Dudarev, G. A. Botton, S. Y. Savrasov, Z. Szotek, W. M. Temmerman, and A. P. Sutton, Phys. Status Solidi A \textbf{166}, 429 (1998).
\bibitem{dorado09} B. Dorado, B. Amadon, M. Freyss, and M. Bertolus, Phys. Rev. B \textbf{79}, 235125 (2009).
\bibitem{hybrid} K. N. Kudin, G. E. Scuseria, and R. L. Martin, Phys. Rev. Lett. \textbf{89}, 266402 (2002);
I. D. Prodan, G. E. Scuseria, and R. L. Martin, Phys. Rev. B \textbf{73}, 045104 (2006);
\emph{ibid} \textbf{76}, 033101 (2007).

\bibitem{geng08b}  H. Y. Geng, Y. Chen, Y. Kaneta, and M. Kinoshita, Phys. Rev. B \textbf{77}, 180101(R) (2008).
\bibitem{geng08c}  H. Y. Geng, Y. Chen, Y. Kaneta, and M. Kinoshita, Appl. Phys. Lett. \textbf{93}, 201903 (2008).

\bibitem{higgs07} J. D. Higgs, W. T. Thompson, B. J. Lewis, and S. C. Vogel, J. Nucl. Mater. \textbf{366}, 297 (2007).

\bibitem{andersson09} D. A. Andersson, J. Lezama, B. P. Uberuaga, C. Deo, and
S. D. Conradson, Phys. Rev. B \textbf{79}, 024110 (2009).

\bibitem{hering52} J. Hering and P. Perio, Bull. Soc. Chim. Fr. \textbf{19}, 351 (1952).
\bibitem{bevan86} D. J. M. Bevan, I. E. Grey, and B. T. M. Willis, J. Solid State Chem. \textbf{61}, 1 (1986).
\bibitem{cooper04} R. I. Cooper and B. T. M. Willis, Acta Crystallogr., Sect. A: Found. Crystallogr. A\textbf{60}, 322 (2004).

\bibitem{yakub06} E. Yakub, C. Ronchi, and I. Iosilevski, J. Phys.: Condens. Matter \textbf{18}, 1227 (2006).

\bibitem{vasp} G. Kresse and J. Furthm{\"u}ller, Comput. Mater. Sci. \textbf{6}, 15 (1996).
\bibitem{kresse96} G. Kresse and J. Furthm{\"u}ller, Phys. Rev. B \textbf{54}, 11169 (1996).
\bibitem{blochl94} P. E. Bl{\"o}chl, Phys. Rev. B \textbf{50}, 17953 (1994).
\bibitem{kresse99} G. Kresse and D. Joubert, Phys. Rev. B \textbf{59}, 1758 (1999).
\bibitem{anisimov91} V. I. Anisimov, J. Zaanen, and O. K. Andersen, Phys. Rev. B \textbf{44}, 943 (1991).
\bibitem{liechtenstein95} A. I. Liechtenstein, V. I. Anisimov, and J. Zaanen, Phys. Rev. B \textbf{52}, R5467 (1995).
\bibitem{dudarev98b} S. L. Dudarev, G. A. Botton, S. Y. Savrasov, C. J. Humphreys, and A. P. Sutton, Phys. Rev. B \textbf{57}, 1505 (1998).
\bibitem{dudarev00} S. L. Dudarev, M. R. Castell, G. A. Botton, S. Y. Savrasov,
C. Muggelberg, G. A. D. Briggs, A. P. Sutton, and D. T. Goddard, Micron \textbf{31}, 363 (2000).
\bibitem{dorado-cmt} B. Dorado, B. Amadon, G. Jomard, M. Freyss, and M. Bertolus, Phys. Rev. B \textbf{84}, 096101 (2011).
\bibitem{pbe96} J. P. Perdew, K. Burke, and M. Ernzerhof, Phys. Rev. Lett. \textbf{77}, 3865 (1996).
\bibitem{soderlind02} P. S{\"o}derlind, Phys. Rev. B \textbf{66}, 085113 (2002).
\bibitem{yun11} Y. Yun, J. Rusz, M. T. Suzuki, and P. M. Oppeneer, Phys. Rev. B \textbf{83}, 075109 (2011).

\bibitem{watanabe09} T. Watanabe, S. G. Srivilliputhur, P. K. Schelling, J. S. Tulenko, S. B. Sinnott, and S. R. Phillpot, J. Am. Ceram. Soc. \textbf{92}, 850 (2009).

\bibitem{geng05} H. Y. Geng, M. H. F. Sluiter, and N. X. Chen, Phys. Rev. B \textbf{72}, 014204 (2005).
\bibitem{moruzzi88} V. L. Moruzzi, J. F. Janak, and K. Schwarz, Phys. Rev. B \textbf{37}, 790 (1988).
\bibitem{serizawa99} H. Serizawa, Y. Arai, M. Takano, and Y. Suzuki, J. Alloys Compd. \textbf{282}, 17 (1999).
\bibitem{serizawa00} H. Serizawa, Y. Arai, and Y. Suzuki, J. Nucl. Mater. \textbf{280}, 99 (2000).

\bibitem{garrido03} F. Garrido, R. M. Ibberson, L. Nowicki, and B. T. M. Willis, J. Nucl. Mater. \textbf{322}, 87 (2003).
\bibitem{nowicki00} L. Nowicki, F. Garrido, A. Turos, and L. Thome, J. Phys. Chem. Solids \textbf{61}, 1789 (2000).
\bibitem{bevan86b} D. J. M. Bevan and S. E. Lawton, Acta Crystallogr., Sect. B: Struct. Sci. \textbf{B42}, 55 (1986).
\bibitem{murray90} A. D. Murray and B. T. M. Willis, J. Solid State Chem. \textbf{84}, 52 (1990).

\bibitem{lam74} D. J. Lam, J. B. Darby, and M. B. Newitt, in \emph{The Actinides, Electronic
Structure and Related Properties}, edited by A. J. Freeman and J. B. Darby (Academic, New York,
1974), Vol. 11, p.\,119.
\bibitem{ellis76} W. P. Ellis, Surf. Sci. \textbf{61}, 37 (1976).
\bibitem{winer86} K. Winer, C. A. Colmenares, R. L. Smith, and F. Wooten, Surf. Sci. \textbf{177}, 484 (1986).
\bibitem{eckle04} M. Eckle and T. Gouder, J. Alloys Compd. \textbf{374}, 261 (2004).
\bibitem{brooks84} M. S. S. Brooks, J. Phys. F: Met. Phys. \textbf{14}, 639 (1984).
\bibitem{petit10} L. Petit, A. Svane, Z. Szotek, W. M. Temmerman, and G. M. Stocks, Phys. Rev. B \textbf{81}, 045108 (2010).

\bibitem{fujino81} T. Fujino, H. Tagawa, and T. Adachi, J. Nucl. Mater. \textbf{97}, 93 (1981).
\bibitem{loopstra70} B. O. Loopstra, J. Appl. Crystallogr. \textbf{3}, 94 (1970).
\bibitem{pillai01} C. G. S. Pillai, A. K. Dua, and P. Raj, J. Nucl. Mater. \textbf{288}, 87 (2001).

\bibitem{weller88} M. T. Weller, P. G. Dickens, and D. J. Penny, Polyhedron, \textbf{7}, 243 (1988).

\bibitem{barett63} C. S. Barett, M. H. Mueller, and R. L. Hittermann, Phys. Rev. B \textbf{129}, 625 (1963).
\bibitem{yoo98} C. S. Yoo, H. Cynn, and P. S{\"o}derlind, Phys. Rev. B \textbf{57}, 10359 (1998).

\bibitem{lidiard66} A. B. Lidiard, J. Nucl. Mater. \textbf{19}, 106 (1966).
\bibitem{matzke87} Hj. Matzke, J. Chem. Soc., Faraday Trans. {2} \textbf{83}, 1121 (1987).

\bibitem{kikuchi51} R. Kikuchi, Phys. Rev. \textbf{81}, 988 (1951).
\bibitem{connolly83} J. W. D. Connolly and A. R. Williams, Phys. Rev. B \textbf{27}, 5169 (1983).
\bibitem{geng05b} H. Y. Geng, N. X. Chen, and M. H. F. Sluiter, Phys. Rev. B \textbf{71}, 012105 (2005).
\bibitem{sluiter90} M. Sluiter, D. de Fontaine, X. Q. Guo, R. Podloucky, and A. J. Freeman, Phys. Rev. B \textbf{42}, 10460 (1990).
\bibitem{sluiter96} M. H. F. Sluiter, Y. Watanabe, D. de Fontaine, and Y. Kawazoe, Phys. Rev. B \textbf{53}, 6137 (1996).
\bibitem{note01} The configurational entropy contributed by independent point defects is
proportional to $-k_{B}\sum_{i}N_{i}\left[n_{i}\ln n_{i}+(1-n_{i})\ln(1-n_{i})\right]$, where
$N_{i}$ is the number of available sites for defects. This term, however, is irrelevant when discussion of EOS
is made for the closed regime only.
%===================











%-------------------

\end{thebibliography}
\end{document}